\documentclass[letter,11pt,openright,makeidx,graphicx,bibtex]{book}
\usepackage{index}
\usepackage{fullpage}

\usepackage{times,amsfonts}
\usepackage{epsf}
\usepackage{wrapfig}
\usepackage{graphicx,times}
\usepackage{graphics}
\usepackage{subfig}
\usepackage{setspace, dropping, section, relsize}
\usepackage{caption2, topcapt}
\usepackage[Benne]{fncychap}
\usepackage{amsmath}
\usepackage[dutch,english]{babel}
\usepackage{amssymb,latexsym,stmaryrd}
\usepackage[innercaption]{sidecap}
\usepackage{natbib, natbibspacing}
\usepackage{multirow}
\usepackage{multicol}
\usepackage{deluxetable}
\usepackage{bigdelim}
\usepackage{bigstrut}
\bibpunct{(}{)}{;}{a}{}{,}
\usepackage[flushleft]{threeparttable}

\usepackage{fancyheadings}
\renewcommand{\chaptermark}[1]%
   {\markboth{\MakeUppercase{\chaptername}\thechapter.\ #1}{}}

\renewcommand{\captionlabeldelim}{--.}
\renewcommand{\captionfont}{\small}
\renewcommand{\captionlabelfont}{\normalsize \bf}

\newcounter{savesection}
  {\setcounter{savesection}{\value{section}}%
   \setcounter{section}{0}%
   \renewcommand\thesection{\thechapter.\Alph{section}}}%
  {\setcounter{section}{\value{savesection}}}
\renewcommand{\chaptermark}[1]{\markboth{chapter \thechapter:~~ #1}{}}

\makeatletter
\long\def\@makecaption#1#2{%
   \vskip 10\p@
   \setbox\@tempboxa\hbox{{\bf#1:} #2}%
   \ifdim \wd\@tempboxa >\hsize
    {\bf #1:} #2\par
     \else
       \hbox to\hsize{\hfil\box\@tempboxa\hfil}%
   \fi}
\makeatother

\def\cm2{${\rm cm^{-2}}$}              
\def\cm3{${\rm cm^{-3}}$}

\def\A18{$A_{\rm 18}$}

\makeindex 
\usepackage{geometry}

\begin{document}
\pagestyle{empty}  

\oddsidemargin -1.2in 
\topmargin -0.8in
\clearpage
\begin{figure}
\end{figure}
\clearpage

\oddsidemargin 0.0in
\topmargin 0.0in
\headheight     0.3in
\headsep 0.1in
\clearpage
\newpage
     
\vspace{10cm}
The spiff new cover was made by Robin J. Allen, who also made my thesis-cover. Thank you Robin.
\clearpage
\newpage

\pagestyle{fancyplain}

\tableofcontents


\chapter{\label{ch:foreword}Foreword}

Sextractor for dummies is not a book in the famous series, I just liked the 
title. Sextractor is not a toy for grownups, it is a extremely usefull and 
versatile astronomical (and perhaps other disiplines) softwaretool. \\
In the course of my PhD I learned a great deal about it by trial and error 
and from Ed Smith and many others. To have somewhere to explain 
every knob and dial is the motivation behind this manual.\footnote{Yes 
my motto is ``Never read a manual as long there are still buttons and 
levers left..." but in sextractor's case you might want to make an exception. 
Besides I'm perfectly happy if you just skim it.} I hope to make it as accurate and 
complete and possibly even complete as possible. I hope it helps with 
whatever project you are working on.\\

There has been quite some time between updates of this user manual. 
I apologize but certain external factors insisted I write a PhD thesis first. 
In the meantime, both the official documentation and the program have 
been updated. I hope I've now caught up sufficiently that this is again a 
useful manual for the beginning and advanced source extractor user.
\footnote{And if you find a problem, inconsistency or spelling erro...please 
drop me a line at \\ holwerda@stsci.edu }

\chapter{\label{ch:introduction}Introduction}

\begin{figure}[htb]
    \centering
\caption{\ref{fig:se}Just playing around with mkobject in iraf, source extractor and photoshop.}
\end{figure}

Source Extractor \footnote{Sometimes abbreviated to 'sex', like the 
executable. If this funny to you a second time you might want to get out 
more.} is used for the automated detection and photometry of sources in 
fits image-files. SE works on scans of photographic plates as well as CCDs.
However I am assuming you are working with ccd data. \\

\begin{center}
\noindent\fbox{\parbox{0.8\textwidth}{If you goal is to get catalogs of all the 
detected objects with reasonably good photometry from a {\it FITS} file of 
processed astronomical imaging data, then SE is the instrument you 
need.\footnote{"This is not the manual you're looking for..." Just because 
I had not cited Star Wars just yet.}}}\\
\end{center}

If you want REALLY good photometry on only a few objects and you know 
where they are, do not use SE. 
I'll discuss how SE works and subsequently how to start it and which 
parameters do what and what the different parameters in the catalogs mean.
Some, but by no means all, strategies for its use are discussed and to 
illustrate some examples from my own experiences will be given.

This started as a bunch of notes of mine and Ed Smith and is evolving slowly 
beyond that. I hope to provide for slightly more insight into SE than the official 
manual for two reasons:
first, all the parameters (both input and output) will be discussed in one document, 
making it slightly more complete than either of the manuals and secondly it has DON'T 
PANIC written in large friendly letters on the cover.

\chapter{\label{ch:procon}Pros and Cons of SE}

As I tried to point out in the introduction and will emphasize occasionally 
SE is not suited for every astronomy project that needs photometry of objects 
in a field. As everything, it has strong and weak points.

The pros of SE are listed in the manual but the most important ones are:
\begin{itemize}
\item[1.] Speed. SE is made to go through data quickly. And if you're trying to beaver through several square degrees of data, speed is GOOD.
\item[2.] The capacity to handle large fits files. SE is coded up so that it'll take it a piece at a time. Again good for the sky-eaters among us.
\item[3.] Works on CCD and scanned photographic plate data. Nice if you happen  to have this kind of data.
\item[4.] Does decent photometry. 
\item[5.] Robust, it'll run with idiotic input.
\item[6.] Controllable, most steps can be influenced by user. 
\item[7.] The possibility to accept user specified flag images or weight images.
\item[8.] Output parameters and the order in which these are listed, are specified by the user.
\item[8.] Output image-files depicting apertures, detections and more.
\item[9.] The possibility to detect sources in one image and do the photometry in another.
\item[10.] There is follow-up software to decomposition of galaxy profiles. Source identification and a first selection can be done with source extractor and either GIM2D or GALFIT can then follow-up.
\end{itemize}

However SE has some drawbacks. It was made for speedy use and in some case accuracy has been sacrificed for speed on purpose. So here is the other side of the coin:

\begin{itemize}
\item[1.] Only as good as its settings. SE is dependent on some of it's setting and these are crucial for the detection and photometry. It will run on just about any set of input parameters but give back output that may be total bogus.
\item[2.] Manuals are outdated and incomplete. This handbook is written as a remedy for that but by a user and not the person who wrote the code.
\footnote{So I'm NOT claiming completeness or correctness.}
\item[3.] Limited accuracy. You'll see this with the geometrical output parameters. These are computed (from moments), NOT fitted (which would be more accurate). 
\item[4.] Classification of objects is of {\it very } limited use. 
\item[5.] Breaks down in crowded fields eventually. 
\item[6.] Corrections of photometry for the 'wings' of object profiles is very rudimentary.
\end{itemize}

But as long as you, the user are aware of these little drawbacks and use the SE input as a fist start to fit with GIM2D or something then all will be well.

\chapter{\label{ch:installation}How to install SE}
\index{installation}

\section{\label{sec:oldinstall}Installing version 2.2}

First you get the most recent version of SE from {\it http://terapix.iap.fr/soft/sextractor/index.html}. Then you unzip and tar the file (UNIX):

\begin{verbatim}
gzip -dc sex_2.2.2.tar.gz | tar xv
\end{verbatim}

That should leave you with a directory in the directory where you did this 
called {\it sextractor2.2.2/} with instructions on how to install in 
'INSTALL'. Basically you go to the 'sextractor2.2.2/source' directory  and type:

\begin{verbatim}
make SEXMACHINE= 'machine type'
\end{verbatim}

where the 'machine type' can be any of the following possibilities:
\begin{verbatim}
aix     (for IBMs RS6000 running AIX)
alpha   (for DEC-ALPHAs with Digital UNIX)
hpux    (for HP/UX systems)
linuxpc (for PCs running LINUX, using gcc)
linuxp2 (for Pentium2/3/4 PCs running LINUX)
linuxk7 (for Athlon PCs running LINUX)
sgi     (for SGI platforms)
solaris (for SUN-Solaris machines)
sunos   (for SUN-OS platforms)
ultrix  (for DEC stations running ULTRIX)
\end{verbatim}

The SE manual is available in postscript format to you in the 
'sextractor2.2.2/doc' directory. Congrats, you have Source Extractor 
available as an executable 'sex' in the 'sextractor2.2.2/source' directory. 
The 'make' file tries to make a shortcut to this executable in your 'home' directory. 
If this fails, try making an alias for the command or simply type the whole path 
{\it /wherever/sextractor2.2.2/source/sex} .

\section{\label{sec:newinstall} Installing version 2.3 and up}

The simplest way to compile this package is:
\begin{itemize}
\item[1.] `cd' to the directory containing the package's source code and type `./configure' to configure the package for your system.
\item[2.] Type `make' to compile the package.
\item[3.] Type `make install' to install the programs and any data files and
     documentation.
\end{itemize}

But if you use linux, the rpm files are also available.

\section{\label{sec:scisoft}Scisoft}

At the European Southern Observatory, a package of all interesting scientific 
software is being kept up to date and easy to install. Source extractor is part 
of this package and since the installation is almost plug-and-do-science (for 
the mac at least...). I highly recommend it, especially the Mac.\footnote{Ok....I've plugged this enough.}

get it here:
\begin{verbatim}
http://www.eso.org/science/scisoft/  (LINUX)
http://www.stecf.org/macosxscisoft/  (Mac OSX)
\end{verbatim}

\chapter{\label{ch:SEworks}How SE works}

The source extractor package works in a series of steps. It determines the
background and whether pixels belong to background or objects. Then it 
splits up the area that is not background into separate objects and determines 
the properties of each object, writing them to a catalog.

The background determination is treated in the official manual and in section \ref{sec:back}
All the pixels above a certain threshold are taken to belong to an object.
 If there is a saddle point in the intensity distribution (there are two peaks in 
 the light distribution distinct enough), the object is split in different entries in 
 the catalogs. Photometry is done on these by dividing up the intensity of the 
 shared pixels. There is an option to "clean " the catalog in order to eliminate 
artifacts caused by bright objects. Afterward, there is a list of objects with a 
series of parameters measured (ellipticity, size etc.). These are classified into 
stars and galaxies (everything non-star) by a neural network. \\

The first steps are controlled by a number of parameters. How to estimate the 
threshold? How much contrast should there be to split an object? However, the 
classification by the neural network only depends indirectly on the parameters 
controlling the first steps. As the network has been trained on ground based data, 
there might be some doubts on the reliability of this classification as one switches 
to other passbands or instruments. So check up on this classification in the case 
of faint or blended objects.\\

Some of the steps in SE have maps associated with them and these can be 
written to fits files, the 'check' images.

Aside from the input image, SE can handle weightimages and flag-images to 
mark the relative importance of pixels or to flag bad ones.

The fitting of the Point-Spread-Function is not yet operational but it was already indicated in the second flow diagram in manual version 2.1.3.

Steps of SE:
\begin{itemize}
\item[1.] Measure the background and its RMS noise (background and RMS maps). (6.2)
\item[2.] Subtract background.
\item[3.] Filter  (convolve with specified profile). (6.3.2)
\item[4.] Find objects (thresholding). (6.3.1)
\item[5.] Deblend detections (break up detection into different objects. (6.3.3)
\item[6.] Measure shapes and positions.
\item[7.] Clean (reconsider detections, accounting for contributions from neighbors) (6.3.4)
\item[8.] Perform photometry.
\item[9.] Classify/index level of fuzziness --> more star-like or galactic?
\item[10.]  Output Catalog and 'Check' Images
\end{itemize}

\begin{figure}[htb]
    \centering
    \caption{SE flow diagram from the first manual. This shows the principal steps done by SE.}
    \label{fig:SEflow1}
\end{figure}

\begin{figure}[htb]
    \centering
    \caption{SE flow diagram from the second manual. This shows clearly the many extra options over a simple run of SE; the possibilities of weight maps, flags, crosscorellation catalogs and many more.}
     \label{fig:SEflow2}
\end{figure}

\chapter{\label{ch:SEuse}Using SE}

Source Extractor can be used in basically three ways: on a single file 
for both detection and photometry, on two files, one for detection, the 
other for photometry and on two files with cross-identification in the 
catalogs.\\

\section{\label{sec:1image}Using SE on one image}

SE needs a series of parameters in order to run and these can be given 
at the command line or in a configuration file. So to run SE on a single 
file with all the necessary parameters in the configuration file one types:\\

sex {\it image} -c configuration\_file.txt\\

If there is no configuration file given, SE will try to read 'default.sex' from the local directory. 
However, the parameter values can be fed to SE on the command line as well:\\

sex {\it image} -c configuration\_file.txt -PARAMETER1 {\it value1} -PARAMETER2 {\it value2}\\

The names of the parameters and their meanings and preferred values are 
discussed below. \\
\begin{center}
\noindent\fbox{\parbox{0.8\textwidth}{NOTE: if you use both a configuration file and command 
line parameter input, the command line input takes precedence over the 
configuration file value.}}\\
\end{center}
\section{\label{sec:2image}Using SE on separate images for detection and photometry}

This may sound weird but the option in SE to find all the objects in one 
image and then apply the apertures and positions found on another image 
can be quite useful. To use {\it image1} for the detection of sources and {\it image2} for the photometry:\\

sex {\it image1,image2} -c configuration\_file.txt\\

Suppose you want different colors or color information from the series of images 
you have in different filters. It is in that case very convenient to use the same 
apertures. So provided the images are well aligned, \footnote{The bright sources 
are on the same pixels, check by loading both in saotng or another display package 
and then 'blink' between them. If you're drizzling or stacking several exposures, 
use the same reference image.} the photometry done is essentially the same objects 
using the same apertures.\footnote{If you want awfully good photometry then it might be good te 
realize that a point spread function correction is dependent on the 
filter used. But keep in mind that SE does not do THAT good photometry to start with.}
The nice part is that the numbering in the catalogs in this dual mode and the numbering 
in the single mode on the first image are the same. It should spare you a lot of rooting 
around in the catalogs if you want to compare fluxes in different bands for instance.
This particular feature is further discussed in the strategy section.

\section{\label{sec:assoc}Crosscorrellating catalogs. (ASSOC)}

There is a third possibility to get information on objects that occur in 
separate images, for instance in overlapping fields in a survey or 
simulated objects in the SE detections. Basically you run SE on one field and 
take the X and Y positions (in pixels) 
from the catalogs and feed them to SE in a second run with a search radius 
and a priority (the brightest association or the nearest? etc.)
And the matches are printed in a new catalog. See the ASSOC parameters in the 
catalog configuration section.


\section{\label{sec:multifits}Multi-extention fits}

There is a new feature in Source extractor. It now supports multiple extension 
fits files. FITS is now the 'standard' filetype\footnote{As with standards, everyone 
seems to have their own.} for astronomical observations and several images 
can be part of a single fits-file. Many new data-products now include weight 
and quality maps of the astronomical image \footnote{The Advanced Camera 
for Surveys on board HST for instance.}. 

\chapter{\label{ch:input}SE Input: the Configuration File}

\begin{center}
{\it
"On two occasions, I have been asked [by members of Parliament], 
'Pray, Mr. Babbage, if you put into the machine wrong figures, 
will the right answers come out?' I am not able to rightly 
apprehend the kind of confusion of ideas that could provoke 
such a question."
}\\
                               -- Charles Babbage
\footnote{Morale: input is inportant...}
\end{center}

As stated in the previous section, SE tries to read the configuration file 
{\it default.sex} or a file can be given on the command line.\footnote{If 
the idea of having {\it .sex} files littering your hard disk is a little too 
randy for you, SE will read in any ASCII text files it is given as a 
configuration file.} The {\it default.sex} can be found in the 
{\it /sextractor-2.4.4/config/} directory. It gives a good set of defaults for SE to use. 
  
The configuration file is good way of remembering which parameters you used 
in running SE and you do not have to reset all the parameters when running 
SE over a series of files. The configuration file is an ASCII file (plain 
text) with the name of the parameters and the value on separate lines. 
A comment line begins with '\#' and ends with the end-of-line.\\

The parameters are listed alphabetically in the manual. I discuss them 
here in topical order. Input parameters for SE can be roughly divided 
into the following categories: image information, background estimation, 
detection, photometry, catalogs and SE running parameters.\\

\section{\label{sec:imageinfo}Image Information}


SE gets the positional information from the FITS header but most of the 
following parameters must be specified. 
\index{GAIN}
GAIN is the ratio of the number of electrons to the number of ADU.
The GAIN is dependent on the type of CCD you're using and the instrument in 
front of it (for instance the WF2 chip on the HST has a gain of 7).

How the number you use here, the effective gain, relates to the instrument 
gain is as follows:\\

\begin{tabular*}{5in}[l]{p{1.7in} p{1.4in} p{1.7in}}
Effective Gain 		   		& Magnitude zeropoint & Type of image \\
\hline
gain $\times$ total exposure time & zeropoint(1 \ sec) & input image is c/s \\
gain 			   	& zeropoint(1 sec) + 2.5 log$_{10}$(exp. time) & sum of N frames \\
N$\times$gain 		& zeropoint(1 sec) + 2.5 log$_{10}$(av. exp. time) & average of N frames \\
2$\times$N$\times$gain/3 		   & zeropoint(1 sec) + 2.5 log$_{10}$(av. exp. time) & median of N frames \\
\hline
\end{tabular*}\\

\begin{center}
\noindent\fbox{\parbox{0.8\textwidth}{NOTE: There are different strategies you can follow \
using the GAIN and the zeropoints, see the SE strategies section.}}\\
\end{center}

\index{MAG GAMMA}
The MAG\_GAMMA is a relic of when this program was applied to photographic 
plates (scans thereof). SE v2.2.2 did not RUN, in my case, without it specified, even 
though it does not use it while processing your nice CCD data. Fortunately this 
seems fixed in a recent update.
If you have the misfortune that your data are still scanned in photographic 
plates, then this is the slope of the response function of the emulsion used
on the plates in question. 

\begin{center}
\noindent\fbox{\parbox{0.8\textwidth}{NOTE: SE (before v2.4.4) will not run without 
MAG\_GAMMA specified.}}\\
\end{center}

\index{DETECT TYPE}
\index{PHOTO}
\index{photographic plates}
\begin{center}
\noindent\fbox{\parbox{0.8\textwidth}{
What does MAG\_GAMMA do? (from the official manual)\\

Photographic photometry In DETECT TYPE PHOTO mode, SExtractor assumes 
that the response of the detector, over the dynamic range of the image, is logarithmic. 
This is generally a good approximation for photographic density on deep exposures. 
Photometric procedures described above remain unchanged, except that for each 
pixel we apply first the transformation
\begin{equation}
I = I_0 10^{D \over \gamma}
\end{equation}
where  ($\gamma$ = MAG\_GAMMA is the contrast index of the emulsion, D the original pixel 
value from the background-subtracted image, and $I_0$ is computed from the magnitude 
zero-point $m_0$ (specified in ):
\begin{equation}
I_0 = {\gamma \over ln10} 10^{-0.4 ~ m_0}
\end{equation}

One advantage of using a density-to-intensity transformation relative to the 
local sky background is that it corrects (to some extent) large-scale inhomogeneities 
in sensitivity.
}}
\end{center}

\index{DETECT TYPE}
The DETECT\_TYPE specifies what type of data SE is handling, scanned photo plates 
or CCD data. Even with DETECT\_TYPE set to CCD, older versions of SE will still need that MAG\_GAMMA.

\index{MAG ZEROPOINT}
MAG\_ZEROPOINT is the zeropoint for the photometric measurements. 
This is again different if you use counts-per-second images as opposed 
to total counts 
images. The counts-per-second images have the zeropoint specified by the 
instrument handbook (depends on filter, instrument and type of ccd used). 
But in the case of a total counts image is the handbook value plus the 
$ \rm 2.5 ~ log_{10}(exposure time)$.

\index{PIXEL SCALE}
PIXEL\_SCALE is again something you hopefully know before you started to run 
SE. Funny enough this is not read from the Fits header. So specify this!
SE needs it {\it only} for the CLASS\_STAR parameter (but still needs it).

\begin{center}
\noindent\fbox{\parbox{0.8\textwidth}{NOTE: New feature in SE (v2.4.4.). 
When this parameter is set to 0, SE uses the World coordinate fits 
information to compute the pixelscale.}}\\
\end{center}

\index{SATUR LEVEL}
SATUR\_LEVEL is the limit for SE to start extrapolating to get the 
photometry. However as soon as you hit something as saturated as that 
you might want to see if there is another way to determine the flux from 
that object. 

\index{SEEING FWHM}
The SEEING\_FWHM (Full Width at Half Maximum) is important for that 
separation between stars and galaxies. Like the PIXEL\_SCALE it is 
only used for the CLASS\_STAR parameter.\footnote{Seeing is the 
blurring of the image as a result of atmospheric disturbances (cirrus 
clouds, turbulence etc.). An estimate of the seeing should be documented 
in either the header or the observation logs. If not, make something up, possibly inspired by that looks like a bright star.}

\begin{tabular*}{\textwidth}[c]{l l p{0.65in} p{2.1in} }
Parameter &	Default &	Type &	Description \\
\hline
\index{FLAG IMAGE}
FLAG\_IMAGE & flag.fits & strings ($n \leq 4$) & File name(s) of the  flag-image(s) . \\
\index{FLAG TYPE}
FLAG\_TYPE & OR & keyword & Combination expression for the method for flags on the same object: \\
 & & OR  & arithmetical OR, \\
 & & AND &  arithmetical AND, \\
 & & MIN &  minimum of all flag values, \\
 & & MAX &  maximum of all flag values, \\
 & & MOST &  most common flag value. \\
\index{GAIN}
GAIN & - & float & Gain  (conversion factor in e"=ADU) used for error estimates of CCD magnitudes.\\

\index{DETECT TYPE}
DETECT\_TYPE & CCD & keyword & Type of device that produced the image: \\
 & & CCD & linear detector like CCDs or NICMOS,\\
 & & PHOTO &  photographic scan. \\

\index{MAG GAMMA }
MAG\_GAMMA & - & float & $\gamma$ of the emulsion (slope of the response function). Takes effect in PHOTO mode only but {\bf NEEDS} to be specified, even for CCD images.\\ 
\index{MAG ZEROPOINT}
MAG\_ZEROPOINT & - & float & Zero-point offset to be applied to magnitudes.\\ 

\index{PIXEL SCALE}
PIXEL\_SCALE & - &  float & Pixel size in arcsec. (for surface brightness parameters, FWHM and star/ galaxy separation only). \\
\index{SATUR LEVEL}
SATUR\_LEVEL & - &  float & Pixel value above which it is considered saturated.\\
\index{SEEING FWHM}
SEEING\_FWHM  & - &  float & FWHM of stellar images in arcsec. This quantity is used only for the neural network star/galaxy separation as expressed in the CLASS\_STAR output.\\
\hline
\end{tabular*}\\

\section{\label{sec:back}Background Estimation}

SE estimates the background of the image as well as the RMS noise in that 
background, mapping both 
\footnote{CHECKIMAGE\_TYPE BACKGROUND and BACKGROUND\_RMS
 if you want to inspect them. See also the checkimage section (section \ref{sec:check}).}. 
SE subtracts the estimated background from the photometry and uses the 
RMS to estimate errors. So the background is important in the rest of the SE run.\\

\index{BACK SIZE}
BACK\_SIZE regulates the estimate. In an area of the BACK\_SIZE, the mean
 and the $\sigma$ of the distribution of pixel values is computed. Then 
the most deviant values are discarded and median and standard deviation
$\sigma$ are computed again. This is repeated until all the remaining 
pixel values are within mean $ \pm 3\sigma$. If $\sigma$ dropped with 
less than 20\% per iteration, the field is considered not crowded.\\

\noindent The value for the background in the area is:
\begin{itemize}
\item the mean in the non-crowded case
\item 2.5 $\times$ median - 1.5 $\times$ mean in the crowded case
\end{itemize}

Both the mean and the median are the ones computed in the last iteration. 
The mean is the average and the median is the average of all the values 
except the most extreme one.\\
The background map is a bi-cubic-spline interpolation over all the area's 
of size BACK\_SIZE, after filtering. So obviously, the choice of 
BACK\_SIZE is very important, too small and the background estimate will 
be partly object flux, too large and small scale variations cannot be 
taken into account. The effect of different BACK\_SIZE values is illustrated 
in figure \ref{fig:back_size}.
\\

\begin{center}
\noindent\fbox{\parbox{0.8\textwidth}{
 NOTE: The BACK\_SIZE parameter determines the background map. 
Estimate the average size of the objects in pixels and make sure the 
BACK\_SIZE is larger than that.}}
\end{center}

But before the fit to the background values is done, the values can be 
smoothed:
\index{BACK FILTERSIZE}
BACK\_FILTERSIZE is the median filter for the background map. 
Effectively, you smooth the background image over this number of meshes
 to get rid of the deviations resulting from bright or extended objects. The effect 
 of different BACK\_FILTERSIZE values is illustrated in figure \ref{fig:back_filtersize}.

\index{BACK TYPE}
BACK\_TYPE is only used if you do not want SE to go off and estimate 
the background but use only one constant value supplied by you in 
\index{BACK VALUE} BACK\_VALUE.

The background computed above is first used in the detection of objects. 
Subsequently this background value can also be used for the photometry. 
But to get accurate background values for the photometry, the background 
can be recomputed in an area centered around the object in question. 
To recompute, set BACKPHOTO\_TYPE to LOCAL and pick an 
BACKPHOTO\_THICK to match your tastes (generally speaking, somewhat
larger than the objects in questions would be a good idea). 

\begin{center}
\noindent\fbox{\parbox{0.8\textwidth}{
NOTE: the RMS as determined from the BACKGROUND\_RMS map 
will be used in more than just the photometry, the thresholds 
for detection and analysis can be dependent on it.}}
\end{center}

\begin{center}
\noindent\fbox{\parbox{0.8\textwidth}{
NOTE: if you want to subtract the background and not have SE do this for you, set BACK\_TYPE to MANUAL and BACK\_VALUE to 0.0,0.0}}
\end{center}

\begin{figure}[htb]
    \centering
    \caption{The HDF-N with the background estimates using different BACK\_SIZE values. Clearly a  BACK\_SIZE smaller than the biggest objects will result in a too much variation in the background estimate. A overlarge BACK\_SIZE will not account for variations in background. Grayscales for the background field are the same. }
    \label{fig:back_size}
\end{figure}

\begin{tabular*}{5in}[l]{l l p{0.65in} p{1.6in} }
Parameter 	 & Default &	Type &	Description \\
\hline

\index{BACK SIZE}
BACK\_SIZE  & - & integers ($n \leq 2$)  &  Size, or Width, Height (in pixels) of a background mesh. \\

\index{BACK FILTERSIZE}
BACK\_FILTERSIZE  & - &  integers ($n \leq 2$)  & Size, or Width, Height (in background meshes) of the background-filtering mask.\\

\index{BACK TYPE}
BACK\_TYPE  & AUTO  & keywords ($n \leq 2$)  &  What background is subtracted from the images:\\
 & & AUTO   &  The internal interpolated background-map.
In the manual it says ``INTERNAL'' here but the keyword is AUTO.\\
 & & MANUAL  &  A user-supplied constant value provided in BACK VALUE.\\ 

\index{BACK VALUE}
BACK\_VALUE & 0.0,0.0  &floats ($n \leq 2$)   & in BACK TYPE MANUAL mode, the constant value to be subtracted from the images. \\

\index{BACKPHOTO THICK}
BACKPHOTO\_THICK & 24  &integer & Thickness (in pixels) of the background LOCAL annulus. \\

\index{BACKPHOTO TYPE}
BACKPHOTO\_TYPE & GLOBAL & keyword & Background used to compute magnitudes:\\
 & & GLOBAL & taken directly from the background map, \\
 & & LOCAL & recomputed in a  rectangular annulus  around the object.\\
\hline
\end{tabular*}\\

\begin{figure}[htb]
    \centering
    \caption{The HDF-N with the background estimates using different BACK\_FILTERSIZE values.
    The mesh of background estimates (BACK\_SIZE = 64) is smoothed over the BACK\_FILTERSIZE before interpolating into the BACKGROUND map. }
    \label{fig:backfiltersize}
\end{figure}

\subsection{\label{sec:weightimages}Weight Images}

\index{Weight Image}
\index{WEIGHT IMAGE}

The individual pixels in the detection image can be given relative 
importance by using a weight for each of them. Different options are available:
the background as determined by SE or an external weight image.
If an external weight image is given, it has to be specified what kind it is; 
a variance map or a rms map or a weight map, from which a variance map 
should be derived. 

The weight for each pixel is derived as follows:

$$ weight = {1 \over variance} = {1 \over rms^2} $$

WEIGHT\_TYPE MAP\_WEIGHT is directly  taken as the weight, 
WEIGHT\_TYPE MAP\_VAR is inverted and WEIGHT\_TYPE MAP\_RMS 
is squared and inverted. 

Reasons for using a weight image are various but to give you an idea:
SE can ignore pieces of the image this way, use a flat-field in it's 
photometry or edited results from a previous run.
\footnote{Please note that I do not specify by what kind of black 
magic you obtained these weight images. They may be the result from a 
previous SE run but a ccd dark-frame might be used. Or perhaps your 
data-reduction scheme will produce a good estimate of the rms. 
To be used with caution!}

Users of the stsdas {\it drizzle} package should check out the appendix on 
drizzle's weight image.

These are the parameters controlling it:\\

\label{WEIGHT_GAIN} 
I can't really fathom why WEIGTH\_GAIN isn't simply an option in 
WEIGHT\_TYPE but it isn't. So there.\\

\label{WEIGHT_IMAGE} 
WEIGHT\_IMAGE is the input parameter where the {\t fits} file is 
given which is to be used as weight map of the type defined by the 
WEIGHT\_TYPE parameter.\\

\label{WEIGHT_TYPE} 
With WEIGHT\_TYPE set on BACKGROUND, the 'checkimage' (output image with 
CHECKIMAGE\_TYPE set on BACKGROUND) of a previous run can be used. 
MAP\_RMS can be for instance derived from known noise characteristics of 
the instrument or given by other programs used in data-reduction 
(for instance the 'drizzle' package for HST data).

\begin{tabular*}{5in}[l]{l l p{0.7in} p{1.9in} }
Parameter &	Default &	Type &	Description \\
\hline
\index{WEIGHT GAIN}
WEIGHT\_GAIN & Y & boolean & If true, weight maps are considered as gain maps. \\

\index{WEIGHT IMAGE}
WEIGHT\_IMAGE & weight.fits & strings      ($n \leq 2$)  & File name of the detection and measurement weightimage , respectively. \\

\index{WEIGHT TYPE}
WEIGHT\_TYPE & NONE & keywords ($n \leq 2$)  &  Weighting scheme (for single image, or detection and measurement images):\\

 & & NONE &  no weighting, \\
\index{BACKGROUND}
 & & BACKGROUND & \\
 & & & variance-map derived from the image itself,\\ 
 & & MAP\_RMS &  variance-map derived from an external RMS-map, \\
 & & MAP\_VAR &   external variance-map, \\
 & & MAP\_WEIGHT & \\
 & & & variance-map derived from an external weight-map,\\

\hline
\end{tabular*}

\begin{center}
\noindent\fbox{\parbox{0.8\textwidth}{
NOTE: WEIGHT\_TYPE set to BACKGROUND does NOT mean that the weight 
image will be used for the background determination.}}
\end{center}

\begin{center}
\noindent\fbox{\parbox{0.8\textwidth}{
This is the description from the SE manual v2.3:
\begin{itemize}
\item[]NONE: No weighting is applied. The related WEIGHT IMAGE and WEIGHT THRESH (see below)parameters are ignored.(quick toggle so you can see if including the weight image has any effect) \\
\item[]BACKGROUND: the science image itself is used to compute internally a variance map (the related WEIGHT IMAGE parameter is ignored). Robust (3 -clipped) variance estimates are first computed within the same background meshes as those described in x??12. The resulting low-resolution variance map is then bicubic-spline-interpolated on the y to produce the actual full-size variance map. A check-image with CHECKIMAGE TYPE MINIBACK RMS can be requested to examine the low-resolution variance map.
\item[]MAP\_RMS: the FITS image speci ed by the WEIGHT\_IMAGE  le name must contain a weightmap
in units of absolute standard deviations (in ADUs per pixel).
\item[]MAP\_VAR: the FITS image speci ed by the WEIGHT\_IMAGE  le name must contain a weightmap
in units of relative variance. A robust scaling to the appropriate absolute level is
then performed by comparing this variance map to an internal, low-resolution, absolute
variance map built from the science image itself.
\item[]MAP\_WEIGHT: the FITS image specified by the WEIGHT\_IMAGE  file name must contain a
weight-map in units of relative weights. The data are converted to variance units (by definition variance / 1=weight), and scaled as for MAP\_VAR. MAP\_WEIGHT is the most commonly
used type of weight-map: a flat-field, for example, is generally a good approximation to a
perfect weight-map.
\end{itemize}
In fact, if you want to work with Weight maps (since you think you have something that might be used as a weigth map), please read section 7 in the official manual.
}}
\end{center}

\clearpage

\section{\label{sec:findobj}Finding and Separating Objects}

SE considers every pixel above a certain threshold (to be specified by YOU
\footnote{yes YOU! You did it to yourself, you and only you and that's what really hurts. (Radiohead)}
, directly or indirectly) to be part of an object. The 'deblending' is the 
part where it figures out which pixels or parts of pixels 
belong to which objects.

\subsection{\label{sec:thresh}Detection; Thresholds}

The threshold parameters indicate the level from which SE should start 
treating pixels as if they were part of objects, determining parameters 
from them.
There are three requirements for a candidate objects:
\begin{itemize}
\item[1.] All the pixels are above the DETECT\_THRESH.
\item[2.] All these pixels are adjacent to each other (either corners or sides in common).
\item[3.] There are more than the minimum a number of 
pixels (specified in DETECT\_MINAREA).
\end{itemize}

\index{ANALYSIS THRESH}
\index{DETECT THRESH}
ANALYSIS\_THRESH is just the threshold for CLASS STAR and FWHM, all the other 
parameters are determined from the DETECT\_THRESH. 

\begin{center}
\noindent\fbox{\parbox{0.8\textwidth}{
NOTE: All the OTHER analysis (photometry and the like) is done with the 
DETECT\_THRESH!}}
\end{center}

They can both be specified in three ways:
\begin{itemize}
\item In Surface Brightness: SBlimit,SBzeropoint e.g. DETECT\_THRESH 23.5, 24
\item In ADU's (set THRESH\_TYPE ABSOLUTE) e.g. DETECT\_THRESH 1.2
\item Relative to background RMS (set THRESH\_TYPE RELATIVE) e.g. DETECT\_THRESH 1.2 (this is the initial setup in the default.sex file.)
\end{itemize}

The  threshold in surface brightness mu ($mag/arcsec^2$ ) needs a calibration 
Zero-point ($mag/arcsec^2$ corresponding to 0 counts). Note that this can/will be
 different from the MAG\_ZEROPOINT value.
\footnote{For the HST instruments, the zeropoints can be obtained using 
CALCPHOT in the stsdas package in IRAF. Remember that these are dependent on 
instrument and filter.}

\begin{center}
\noindent\fbox{\parbox{0.8\textwidth}{
NOTE: This whole SB threshold stuff seems quite the thing until you realize SE just does this:
$$\rm thresh = 10^{- {SB_{limit} - SB_{zeropoint} \over 2.5}} $$
}}
\end{center}


With THRESH\_TYPE set to ABSOLUTE, the threshold is set to the same 
number of ADUs across the image. If THRESH\_TYPE set to RELATIVE 
(the default), the the threshold is that number of background RMS standard 
deviations above the background value. This is nice and flexible but 
sensitive to the background estimation!

The \index{DETECT MINAREA} DETECT\_MINAREA is the minimum 
number of pixels above the threshold required to be considered an object.

\begin{figure}[htb]
    \centering
    \caption{An object is defined as a series of bordering pixels above the threshold. Bordering pixels to the gray one are 1 through 8. Another pixels needs to be above the threshold and share a corner or a side with another pixel in order to considered part of the same object (unless they are deblended into different objects).}
    \label{fig:minarea}
\end{figure}
\clearpage

\begin{tabular*}{5in}[c]{l l p{0.8in} p{1.5in} }
Parameter &	Default &	Type &	Description \\
\hline

\index{THRESH TYPE}
THRESH\_TYPE & RELATIVE & keywords ($n \leq 2$) & Meaning of the DETECT THRESH and ANALYSIS\_THRESH parameters : \\
 & & RELATIVE & scaling factor to the background RMS, \\
 & & ABSOLUTE & absolute level (in ADUs or in surface brightness). \\

\index{ANALYSIS THRESH}
ANALYSIS\_THRESH & - 	& floats ($n \leq 2$) &  Threshold (in surface brightness) at which CLASS STAR and FWHM operate.
1 argument: relative to Background RMS. 
2 arguments: mu ($mag/arcsec"^2$ ), Zero-point (mag).\\

\index{DETECT THRESH}
DETECT\_THRESH  & - &    floats ($n \leq 2$) & Detection threshold. 
1 argument: (ADUs or relative to Background RMS, see THRESH TYPE). 
2 arguments: R (mag.arcsec" 2 ), Zero-point (mag). \\

\index{DETECT MINAREA}
DETECT\_MINAREA  & - &  integer & Minimum number of pixels above threshold triggering detection. \\

\hline
\end{tabular*}\\

\subsection{\label{sec:filter}Filtering}

Before the detection of pixels above the threshold, there is the 
option of applying a filter. This filter essentially smooths the image.
\footnote{The photometry is still being done on the original image, 
don't worry.}

There are some advantages to applying a filter before detection. It may help detect faint, extended objects

However it may not be so helpful if your data are very crowded. 
There are four types of filter to be found in the {\it 
./sextractor2.2.2/config/} directory; Gaussian, Mexican hat, 
tophat and blokfunction of various sizes, all normalized.

This is what the helpful README in this directory tells us:

\begin{tabular*}{5in}[l]{l p{3.8in} }
Name		& Description \\
\hline
default.conv	& a small pyramidal function (fast)\\
gauss*.conv	& a set of Gaussian functions, for seeing FWHMs between 1.5 and 5 pixels (best for faint object detection). \\

tophat*.conv	& a set of "top-hat" functions. Use them to detect extended, low-surface brightness objects, with a very low THRESHOLD.\\
mexhat*.conv	& "wavelets", producing a passband-filtering of the image, tuned to seeing FWHMs between 1.5 and 5 pixels. Useful in very crowded star fields, or in the vicinity of a nebula. WARNING: may need a high THRESHOLD!!\\

block\_3x3.conv	& a small "block" function (for rebinned images like those of the DeNIS survey).\\
\hline
\end{tabular*}\\

The naming convention seems to be: name\_seeingFWHM\_size.conv. Both the seeing FWHM and the size are in pixels.
So depending on what you are after, choose a filter and approximately 
your seeing.

\begin{center}
\noindent\fbox{\parbox{0.8\textwidth}{
NOTE: filter choice and threshold choice are interdependent!
}}
\end{center}

\begin{tabular*}{5in}[l]{l l p{0.7in} p{2.0in} }
Parameter &	Default &	Type &	Description \\
\hline

\index{FILTER}
FILTER  & - &   boolean & If true, filtering is applied to the data before extraction. \\

\index{FILTER NAME}
FILTER\_NAME   & - &  string & Name and path of the file containing the filter definition. \\

\index{FILTER THRESH  }
FILTER\_THRESH  & - & floats ($n \leq 2$)  & Lower and higher thresholds (in back-ground standard deviations) for a pix-el to be considered in filtering (used for retina-filtering only). \\
\hline
\end{tabular*}\\

\subsection{\label{sec:deblend}Deblending: separating into different objects}

\index{Deblending}
\index{DETECT THRESH}
\index{DETECT NTHRESH}
\index{DETECT MINCONT}
Deblending is the part of SE where a decision is made whether or not 
a group of adjacent pixels above DETECT\_THRESH is a single object or not.
Suppose there is a little island of adjacent pixels above the threshold. 
It is an object or maybe several really close next to each other.
So how does SE cut this up into different objects? First it defines 
a number of levels between the threshold and the maximum count in 
the object. This is set by the DEBLEND\_NTHRESH parameter. The levels 
are spaced exponentially.

SE then constructs a 'tree' of the objects, branching every time there 
are pixels above a threshold separated by pixels below it (see figure).
A branch is considered a different object provided:

\begin{itemize}
\item[1.] The number of counts in the branch (A in the figure) is above a certain
fraction of the total count in the entire 'island'.
\item[2.] There is at least one other branch (yep B!) above the same level that is also above this fraction.
\end{itemize}

\noindent The fraction is defined in DEBLEND\_MINCONT. 
\footnote{Note that 0.01 is 1\% of the flux when defining this.}

\begin{figure}[htb]
    \centering
    \caption{Illustration of DEBLEND\_NTHRESH and DEBLEND\_MINCONT.}
    \label{fig:seg}
\end{figure}

\begin{tabular*}{5in}[l]{l l l p{1.9in} }
Parameter &	Default &	Type &	Description \\
\hline

\index{DEBLEND MINCONT}
DEBLEND\_MINCONT & - &  float & Minimum contrast parameter for deblending. \\ 

\index{DEBLEND NTHRESH}
DEBLEND\_NTHRESH & - &  integer & Number of deblending sub-thresholds.\\  

\hline
\end{tabular*}\\

\subsection{\label{sec:clean}Cleaning}

There is the option to 'clean' the list of objects of artifacts due to 
bright objects (set CLEAN to YES). 
All the detections are checked to see if they would 
have been detected (i.e. exceeded the threshold etc.) if their neighbors 
were not there. To do this, the contributions of the neighboring 
objects has to be computed. An estimate is made from a moffat light 
profile. The Moffat profile is scaled and stretched to fit the neighbour's 
profile. The contribution to the object from the wings of the Moffat profile 
is then subtracted.

\index{Moffat profile}
The Moffat profile looks like this:

$$ {I(r) \over I(0)} = {1 \over (1 + k \times r^2)^{\beta}}$$

\noindent The CLEAN\_PARAM is the $\beta$ parameter in the above formula.\\

\begin{center}
\noindent\fbox{\parbox{0.8\textwidth}{
NOTE: Decreasing CLEAN\_PARAM yields brighter wings and more aggressive cleaning.
}}
\end{center}

\noindent The value for the CLEAN\_PARAM should be between 0.1 and 10.\\

\begin{center}
\noindent\fbox{\parbox{0.8\textwidth}{
NOTE: In earlier versions of SE\footnote{Those that appeared shortly after the invention of the wheel...}, the Moffet profile was a Gaussian and 
the CLEAN\_PARAM the stretch factor for the FWHM.
}}
\end{center}

\noindent Cleaning would be more aggressive with a higher CLEAN\_PARAM.
This version of cleaning is explained in the original manual. 

\begin{tabular*}{5in}[l]{l l p{0.7in} p{2.0in} }
Parameter & Default & Type &	Description \\
\hline

\index{CLEAN}
CLEAN & -  & boolean & If true, a  cleaning  of the catalog is done before being written to disk. \\

\index{CLEAN PARAM}
CLEAN\_PARAM & -  & float & Efficiency of cleaning.\\ 
\hline
\end{tabular*}\\

\section{\label{sec:photin}Influencing Photometry}

After deblending the objects, SE performs astrometry (where stuff is), 
photometry (how bright stuff is) and geometric parameters (how stuff looks 
like). The astrometry cannot be influenced via input parameters 
(you just specify what kind of positions you want, easy). The photometry had a 
few input parameters associated with it; what to do with overlapping pixels, 
what is the zeropoint and how to apply apertures. To understand these, the 
AUTO photometry of SE has to be explained.
The geometric parameters are mainly associated with the Kron (AUTO) photometry.

The GAIN and MAG\_ZEROPOINT, have been discussed at the image 
characteristics section. Of course they are needed for the photometry.
The GAIN to convert counts to flux and the MAG\_ZEROPOINT for the 
calibration of the magnitude scale. Also the \index{BACKPHOTO THICK}
BACKPHOTO\_THICK \index{BACKPHOTO TYPE} and BACKPHOTO\_TYPE give you influence 
on the way the background subtracted from the photometry is estimated
(see section \ref{sec:back}). 

There are five different approaches in SE's photometry; isophotal, 
isophotal-corrected, automatic, best estimate and aperture.
\footnote{SE v2.4.4 introduces a sixth, an 'automatic' photmetry with the Petrosian radius.}

\subsection{\label{sec:iso}ISO}

\index{ISO}
In the above you defined above what threshold SE should do it's 
photometry, with the estimated background as zeropoint. 
The pixels above this threshold constitute an isophotal area. The 
flux or magnitude determined from this (counts in pixels above 
threshold minus the background) is the {\it isophot } flux/magnitude.
Apart from the threshold (DETECT\_THRESH) and the background 
estimation, there is nothing to influence here.

\subsection{\label{sec:isocor}ISOCOR}

\index{ISOCOR}
In real life however, objects rarely have all their flux within 
neat boundaries, some of the flux is in the ``wings'' of the profile. 
SE can do a crude correction for that, assuming a symmetric Gaussian 
profile for the object. This would be the {\it isophot-corrected} 
flux/magnitude. There is no parameter for you to influence this estimate.


\begin{center}
\noindent\fbox{\parbox{0.8\textwidth}{

{\bf How ISOCOR works (from the v2.3 manual)}\\

Corrected isophotal magnitudes (MAG ISOCOR) is a quick-and-dirty way for 
retrieving the fraction of flux in the wings of the object missed in the isophotal 
magnitudes. 
The latest version of the SE manual (v2.3) does not rate this as a very good 
correction, these values have been kept as an output option for compatibility 
with SE v2.x and  SE v1\footnote{Plus, having expended all the effort to code it up, 
why tear it out?}. 
The assumtion is that the profiles of the objects are Gaussian.
The fraction of the total flux enclosed within a particular isophote reads 
compared to the total flux\footnote{There is a reference to a Maddox et al (1990) 
for this correction in the manual. I have not found this paper. }:

\begin{equation}
1- {1\over \eta} ~ ln(1-\eta) = {A t \over I_{iso}}
\end{equation}

where A is the area and t the threshold related to this isophote. 
The relation can not be inverted analytically, but a good approximation to 
($\rm error ~ < ~ 10^2 ~ for ~ \eta ~ > ~ 0.4$) can be done with the second-order polynomial of t:

\begin{equation}
\eta \approx 1 - 0.1961 {A.t \over A_{iso}} - 0.7512 \left( {A.t \over I_{iso} }\right)^2
\end{equation}

A total magnitude $m_{tot}$ estimate is then $\rm m_{tot} = m_{iso} + 2.5 log(\eta)$
Clearly this first correction works best with stars (and maybe starclusters?) as these are 
the most Gaussian-like, ok with disk galaxies and not-so-ok with ellipticals as 
these have broader wings. 
}}
\end{center}

\noindent Fixed-aperture magnitudes (MAG\_APER) estimate the flux above the background 
within a circular aperture. The diameter of the aperture in pixels\\ (PHOTOM\_APERTURES) 
is supplied by the user (in fact it does not need to be an integer since each normal 
pixel is subdivided in $5 \times 5$ sub-pixels before measuring the flux within the aperture). 
If MAG\_APER is provided as a vector MAG\_APER[n], at least n apertures must be 
specified with PHOTOM\_APERTURES.

\subsection{\label{sec:auto}AUTO}

\index{AUTO}
SE uses a flexible elliptical aperture around every detected object
and measures all the flux inside that, described in \cite{Kron1980}. 
There are two parameters regulating the elliptical apertures: 
PHOT\_AUTOPARAMS and PHOT\_AUTOAPERS.

\noindent The characteristic radius for the ellipse is:

\begin{equation}
 r_1 = {\Sigma r I(r) \over \Sigma I(r)}
\end{equation}

\noindent Also known as the Kron radius (see section \ref{sec:R1}). 
From the objects second order moments, the ellipticity $\eta$ and 
position angle $\theta$ are computed. The major and minor axes of 
the elliptical aperture are computed to be $k \dot r_1/\epsilon$ and 
$\epsilon \dot k \dot r_1$ respectively.
PHOT\_AUTOPARAMS influences directly the estimation. The first is 
k factor mentioned above and the second is the minimum radius for a 
Kron ellipse.\\ PHOT\_AUTOAPERS are the minimum aperture diameters for 
the Kron photometry, estimation and measurement. These are circular! 
These are used in case the radius of the Kron aperture goes below 
the $R_{min}$ specified in PHOT\_AUTOPARAMS. 
The values in the default.sex are probably best for most applications.
A good way to check the setting on PHOT\_AUTOPARAMS is to generate an
'APERTURES' check-image and see if the Kron apertures overlap too much with
those of neighboring objects.\footnote{Note that the elliptical apertures in figure \ref{fig:se} (Source Extractor written in sources? remember? Did you skim over it? you bastard!) are actual Kron apertures for the 'objects'. I cheated however and set PHOT\_AUTOPARAMS to 1.0 so the Kron radii would not be so big.}

\begin{tabular*}{\textwidth}[l]{l l p{0.8in} p{1.4in} }
Parameter &	Default &	Type &	Description \\
\hline
\index{PHOT AUTOPARAMS}
PHOT\_AUTOPARAMS & - &  floats (n = 2) & MAG AUTO controls: scaling parameter k of the 1st order moment, and minimum Rmin (in units of A and B). \\
\index{PHOT AUTOPERS}
PHOT\_AUTOAPERS & 0.0,0.0 & floats (n = 2) & MAG AUTO minimum (circular) aperture diameters: estimation disk, and measurement disk. \\
\hline
\end{tabular*}\\

\begin{center}
\noindent\fbox{\parbox{0.8\textwidth}{
{\bf Neat trick:} The Kron radius was introduced as a accurate flexible 
aperture that would capture most of the flux from an object. There is 
relation between the Kron flux-measurement and the total flux from an 
object given in \cite{Graham05a} depending on the Sersic profile of the 
galaxy. See for the enclosed fraction table \ref{table:kronradii}. 
}}
\end{center}

\begin{center}
\noindent\fbox{\parbox{0.8\textwidth}{
NOTE: In the 2.4.4 version of source extractor provides an alternative very 
similar to the AUTO photometry, the PETRO photometry parameters, determined 
within an aperture defined by the Petrosian radius.
}}
\end{center}

\begin{figure}[htb]
    \centering
    \caption{An example of the different choices of the Kron factor (k) in PHOT\_AUTOPARAMS.
    As the Kron aperture grows, it encloses more flux from other objects. Note how the checkimage set to APERTURES is a good check against this. Dodgy Kron apertures are dashed.}
    \label{fig:Kronradii}
\end{figure}

\subsection{\label{sec:best}BEST}

\index{BEST}
With all this flexibility, you'd expect the Kron or AUTO photometry 
to be the best. However it can be influenced by nearby sources. 
Therefore there is a fourth option, MAG\_BEST. This is usually equal 
to AUTO photometry but if the contribution of other sources exceeds 
10\%, it is ISOCOR. 

\begin{center}
\noindent\fbox{\parbox{0.8\textwidth}{
NOTE: the characterization 'BEST' is a bit misleading. I was advised 
from several sides NOT to use this photometry. Both AUTO and ISOCOR 
are at least consistent.
}}
\end{center}

\subsection{\label{sec:aper}Circular Apertures}

\index{PHOT APERTURES}
\index{APER}
\index{circular apertures}
Then there is the user specified apertures. The aperture se 
{\it diameters} are specified in pixels in the PHOT\_APERTURES.
The fluxes, magnitudes and errors in these are specified in
FLUX\_APER,MAG\_APER, FLUXERR\_APER and MAGERR\_APER. 
Please note that upon inclusion of N apertures, these parameters are N 
columns each in the outputcatalog, corresponding to each specified aperture.  

The calibration documentation for the instrument used should give you a photometric 
correction for a certain circular aperture and a point source. It tells you the relation 
between the flux captured within the aperture and all the flux from a star. 
Since part of the flux has been smeared out of the aperture by the PSF, you need to correct for this. 
If you have an extended object, the game is different. Fortunately, there are other apertures\footnote{The Kron and Petrosian ones...see sections \ref{sec:R1} and \ref{secRp}.} for this and even a correction for those to account for all the flux (but that also depends on the light profile of the galaxy). 

\index{PHOT FLUXFRAC}
\index{FLUX RADIUS}
One of the geometric output-parameters is the half-light radius. 
The fraction of total light within this radius is specified in the 
PHOT\_FLUXFRAC parameter.

\begin{figure}[htb]
    \centering
    \caption{Illustration of the different apertures possible: ISO, ISOCOR, AUTO and APER (user specified in PHOT\_APERTURES}
    \label{fig:aper}
\end{figure}

\section{\label{sec:radinput}Typical Radii}
There are several output option to describe the typical size of an object. 
Often they also define an aperture that is meant to capture an extended 
object like a galaxy.
These are the input parameters that are important to them. The FLUX\_RADIUS 
has been an output parameter for some time, the KRON and PETROSIAN 
radii appeared in version 2.4.4 of SE.

\subsection{\label{sec:inRe}Effective Radii}

\index{half-light radius}
\index{FLUX RADIUS}
\index{PHOT FLUXFRAC}
SE has the option to put out radii containing a certain fraction of the light.
(outputparameter FLUX\_RADIUS) The default is 0.5 (the half light radius).
PHOT\_FLUXFRAC 0.2,0.5,0.9 will give three radii containing 20\%, 50\% and 
90\% of the light respectively. The effective radius output is discussed in section 
\ref{sec:Re}.

\begin{tabular*}{5in}[l]{l l p{0.9in} p{1.4in} }
Parameter &	Default &	Type &	Description \\
\hline
\index{PHOT FLUXFRAC}
PHOT\_FLUXFRAC & 0.5 & floats ($n \leq 32$) & Fraction of FLUX AUTO defining each element of the FLUX RADIUS vector. \\
\hline
\end{tabular*}

\subsection{\label{sec:inR1} Kron Radius}

The Kron radius is the typical size of the aperture already described in the AUTO 
photometry section (section \ref{sec:auto}).
The whole photometry process is controlled by the PHOT\_AUTOPARAMS and 
the definition of the Kron radius is given in section \ref{sec:R1}. \footnote{Leisurely background 
reading on the subject does not excist of course but there is \cite{Kron80} who proposed the radius for accurate galaxy photometry and more recently (by about 25 years) \cite{Graham05a} who gives a very nice overview of the whole photometry within a radius relation. Get it from astro-ph where it's free though. You're probably a Graduate student and poor.}

\index{PHOT AUTOPARAMS}
\begin{tabular*}{5in}[l]{l l p{0.7in} p{1.4in} }
Parameter &	Default &	Type &	Description \\
\hline
PHOT\_AUTOPARAMS &	2.5, 1.5	& floats (n=2) & MAG\_AUTO parameters: (Kron\_fact),(min\_radius) \\
\index{PHOT AUTOAPERS}
PHOT\_AUTOAPERS & 0.0,0.0 & floats (n = 2) & MAG AUTO minimum (circular) aperture diameters: estimation disk, and measurement disk. \\
\hline
\end{tabular*}\\

\noindent The Kron\_fact is the numbe of Kron radii the aperture is set at and the 
min\_rad is the minimum Kron radius for which this is done. Otherwise 
the minimum aperture specified in PHOT\_AUTOAPERS is used for the photometry.

\subsection{\label{sec:inRp} Petrosian Radius}

\index{PHOT PETROSIAN}
\index{Petrosian radius}

The Petrosian radius is another defined radius for photometry and this 
parameter has also only one input parameter associated with it, 
PHOT\_PETROPARAMS. There is however a defining parameter for the 
Petrosian parameter and this is the ratio $\eta$. Mostly this is set to 0.2 
but occasionally it is set to 0.5. Unfortunately, it is impossible to change 
this at this time. See for the complete expression of the Petrosian radius 
section \ref{sec:Rp}.

\begin{tabular*}{5in}[l]{l l p{0.7in} p{1.4in} }
Parameter &	Default &	Type &	Description \\
\hline
PHOT\_PETROPARAMS & 2.0, 3.5       & floats (n=2) & MAG\_PETRO parameters: (Petrosian\_fact),(min\_radius) \\
\hline
\end{tabular*}

\noindent The Petrosian\_fact is the number of Petrosian radii the aperture 
is set at and the min\_rad is the minimum Petrosian radius for which this is 
done. Otherwise the minimum radius is used for the photometry.\footnote{If 
you are having deja-vu all over again, it's because I copies this directly from 
the section above...which I'm sure is how the SE code was made as well.}

\begin{figure}[htb]
    \centering
    \caption{In the checkimage there are two radii visible. Good to know: as soon as you specify either AUTO (Kron) parameters or PETRO parameters for the output catalog, these apertures will be drawn in the checkimage (APERTURES). The Petrosian is usually the outer radius.}
    \label{fig:Kron_Petroradii}
\end{figure}

\index{Kron Radius}
\index{CHECK IMAGE}
\index{AUTO}
\index{KRON RADIUS}
\begin{center}
\noindent\fbox{\parbox{0.8\textwidth}{
NOTE: The output parameters asked for and the checkimage with the apertures (checkimage\_type = APERTURES) influence each other. If no parameter depending on the Kron radius (all the AUTO photometry and KRON\_RADIUS) is asked for in the parameter file, the Kron radius is NOT drawn in the ckeckimage.
}}
\end{center}

\clearpage

\subsection{\label{sec:masking}Masking Overlapping Objects}

\index{MASK TYPE}
\index{bad pixels}
Now what if there are two objects overlapping each other? How to account 
for the overlapping pixels? This is handled by the MASK\_TYPE parameter.
NONE means that the counts in the overlap are simply added to the objects 
total. BLANK sets the overlapping pixels to zero. CORRECT, the default, 
replaces them with their counterparts symmetric to the objects' center.
Best if you leave it at default. I'm just mentioning it out of completeness
\footnote{NONE might be useful to get the total flux from a large 
extended underlying structure with bright patches and no foreground stars.}

\begin{tabular*}{5in}[l]{l l p{0.7in} p{1.4in} }

Parameter &	Default &	Type &	Description \\
\hline
\index{MASK TYPE}
MASK\_TYPE & CORRECT & keyword & Method of  masking  of neighbors for photometry:\\
 & &  NONE &  no masking, \\ 
 & & BLANK &  put detected pixels belonging to neighbors to zero, \\ 
 & & CORRECT  & replace by values of pixels symmetric with respect to the source center. \\ 



\hline
\end{tabular*}\\

\section{\label{sec:serun}SE Running}

These inputparameter govern the way SE runs, if it should heed flags, how 
it should heed those, if and what to put in an outputimage, how much it 
should comment and how much memory it should use.

\subsection{\label{sec:flagsin}Flags}

\begin{figure}[htb]
    \centering
    \caption{Well sometimes flags {\bf are} important. Cartoon by S. Bateman, used without any permission whatsoever but I don't make money off this anyway.}
    \label{fig:check}
\end{figure}

If pixels in your image should be flagged as unreliable or other, SE can use 
a flag image for this purpose. This is well described (I think) in the official 
manual so I copies that section in section \ref{sec:flagsout}. The internal 
flags of SE are described in \ref{sec:inflags}. If however you have some 
kind of quality image (such as a Drizzle weight image or a coverage map 
coming out of the MOPEX pipeline), then you could conceivably convert 
this to a flag image to be fed to SE here. SE will then combine your flags 
(from the flagimage\footnote{Specified in FLAG\_IMAGE...ta-dah!}) with 
it's own internal flags\footnote{The ones that state that Timmy...I mean 
your object is too close to the edge etc etc...}.
It also has different ways to combine the flags (specified in FLAG\_TYPE). 
If your flags are in ascending order of awfulness (flag = 1 means okay but 
with a bad pixel, flag=100 means you made this part of the image up...) 
then you could go for MAX or OR option

\index{Flags}

\begin{tabular*}{5in}[l]{l l p{0.7in} p{2in} }

Parameter &	Default &	Type &	Description \\
\hline

\index{FITS UNSIGNED}
FITS\_UNSIGNED & N & boolean & Force 16-bit FITS input data to be interpreted as unsigned integers. \\

\index{FLAGIMAGE}
FLAG\_IMAGE & flag.fits & strings ($n \leq 4$) & File name(s) of the  flagimage(s) . \\

\index{FLAG TYPE}
FLAG\_TYPE & OR & keyword & Combination method for flags on the same object: \\

 & & OR  & arithmetical OR, \\
 & & AND &  arithmetical AND, \\
 & & MIN &  minimum of all flag values, \\
 & & MAX &  maximum of all flag values, \\
 & & MOST &  most common flag value. \\
\hline
\end{tabular*}\\

\subsection{\label{sec:interpol}Interpolation}

If the data for pixels is missing, SE can interpolate. These parameters 
regulate the interpolation. Best kept at default. For some, the x and y gaps allowed are a bit wide (16 pixels after all, it is almost an entire object...).
On the other hand, it allowd Sextractor to give you catalogs despite bad columns. Therefore do not set to zero.

\begin{tabular*}{5in}[l]{l p{0.4in} p{0.7in} p{1.9in} }

Parameter &	Default &	Type &	Description \\
\hline

\index{INTERP MAXXLAG}
INTERP\_MAXXLAG & 16 & integers  ($n \leq 2$)  & Maximum x gap (in pixels) allowed in interpolating the input image(s). \\

\index{INTERP MAXYLAG}
INTERP\_MAXYLAG & 16 & integers ($n \leq 2$)  & Maximum y gap (in pixels) allowed in interpolating the input image(s).\\

\index{INTERP TYPE}
INTERP\_TYPE & ALL & keywords ($n \leq 2$) & Interpolation method from the variance-map(s) (or weight-map(s)):\\
 & & NONE  & no interpolation, \\
 & & VAR\_ONLY &  interpolate only the variance-map (detection threshold), \\
 & & ALL &  interpolate both the variance-map and the image itself. \\

\hline
\end{tabular*}\\

\subsection{\label{sec:memory}Memory Use}

These are the parameters regulating the memory use of SE. To be honest, 
they are best kept at the values in the {\it ./sextractor2.2.2/config/default.sex} file. 
\footnote{SE was programmed to do large images, even with limited memory 
and computing power. The only reason to change these defaults is when you 
get something of a stack overflow.}
If you have the FLAGS output, then you can check if there were any memory problems in the SE run.
If so, this can be very bad for the completeness of your catalog (it isn't probably).
In that case you might want to fiddle with the MEMORY\_BUFSIZE and rerun.

\begin{tabular*}{5in}[l]{l p{0.4in} p{0.6in} p{1.9in} }

Parameter &	Default &	Type &	Description \\
\hline

\index{MEMORY BUFSIZE}
MEMORY\_BUFSIZE & - &  integer & Number of scanlines in the imagebuffer. 
Multiply by 4 the frame width to get equivalent memory space in bytes. \\

\index{MEMORY OBJSTACK}
MEMORY\_OBJSTACK & - &  integer & Maximum number of objects that the 
objectstack can contain. Multiply by 300 to get equivalent memory space in bytes. \\

\index{MEMORY PIXSTACK}
MEMORY\_PIXSTACK & - &  integer & Maximum number of pixels that the 
pixel-stack can contain. Multiply by 16 to 32 to get equivalent memory 
space in bytes.\\ 
\hline
\end{tabular*}

\subsection{\label{sec:nnw}Neural Network}

There is to date only one neural network file and it's in 
the same directory ({\it default.nnw}) as the other config files. 
Use this one. Don't edit it. I think the original idea was to have 
specialized neural network files for different types of instruments 
but it turns out it's much easier to run something on a SE catalog.

\begin{tabular*}{5in}[l]{l p{0.4in} p{0.6in} p{1.9in} }
Parameter &	Default &	Type &	Description \\
\hline
\index{STARNNW NAME}
STARNNW\_NAME & - &  string & Name of the file containing the neural 
network weights for star/galaxy separation. \\
\hline
\end{tabular*}

\subsection{\label{sec:verbose}Comments }

The VERBOSE\_TYPE parameter regulates the amount of comments 
printed on the command line. It could possibly be instructive to run it 
with FULL once in a while. the descriptions are not very helpful but then 
again you only want to use QUIET if SE is part of some kind of pipeline 
and FULL if something is off and you can't figure out what. 

\begin{tabular*}{5in}[l]{l p{0.4in} p{0.6in} p{1.9in} }
Parameter &	Default &	Type &	Description \\
\hline
\index{VERBOSE TYPE}
VERBOSE\_TYPE & NORMAL & \\
 & & keyword & How much SExtractor comments its operations:\\
 & & QUIET & run silently, \\
 & & NORMAL &  display warnings and limited info concerning the work in progress,\\
 & & EXTRA\_WARNINGS & \\
 & & & like NORMAL, plus a few more warnings if necessary, \\
 & & FULL &  display a more complete information and the principal 
 parameters of all the objects extracted. \\
\hline
\end{tabular*}

\section{\label{sec:inoutput}SE output settings}

SE has two types of output. The catalogs with a whole range of
characteristics of each of the detected objects and outputs
which allow you to compare SE estimates of background, apertures and 
objects with the real data.

\subsection{\label{sec:cat}Catalog}

The catalog is what you are running SE for! So in CATALOG\_NAME , 
you specify the name of the output catalog. Again it's probably a 
good idea to start straight away with a naming convention. 
(a .cat extension for instance!)

The CATALOG\_TYPE enables you to specify the type of outputcatalog. 
Personally I prefer the ASCII\_HEAD, as it allows me to read it in 
just about anywhere and still tells me which parameters are listed.
The nice thing about the fits catalog is that all the input parameter
settings are saved in the header. 

\begin{center}
\noindent\fbox{\parbox{0.8\textwidth}{
NOTE: the fits option can't handle array outputinformation such as 
MAG\_APER,FLUX\_RADIUS if more than one value!
}}
\end{center}

\footnote{I have this on Ed's authority. I have NO idea how to display a fits table. Since there have been quite a number of updates, I think this {\it may} be fixed by now. Not very helpful I know. }

The ASCII\_SKYCAT option for instance does not list all the parameters before the actual catalog (like ASCII\_HEAD) but puts the name of the output parameter on top of the column in question. 

And which parameters to list is specified in the file  given to 
PARAMETERS\_NAME. 

\begin{tabular*}{5in}[l]{l p{0.4in} p{0.5in} p{2.0in} }
Parameter & Default & Type &	Description \\
\hline

\index{CATALOG NAME}
CATALOG\_NAME & -  & string & Name of the output catalog. If the name ``STDOUT'' is given and CATALOG TYPE is set to ASCII, ASCII HEAD, or ASCII SKYCAT, the catalog will be piped to the standard output (stdout) \\

\index{CATALOG TYPE}
CATALOG\_TYPE & -   &  keyword & Format of output catalog: \\

 &  & ASCII   & ASCII table; the simplest, but space and time consuming,  \\
 &  & ASCII\_HEAD  & \\
 &  & & as ASCII, preceded by a header containing information about the content,  \\
 &  & ASCII\_SKYCAT & \\
 &  & &  SkyCat ASCII format (WCS coordinates required),  \\
 &  & FITS\_1.0  &  FITS format as in SExtractor 1,  \\
 &  & FITS\_LDAC & \\
 &  &  & FITS ``LDAC''  format (the original image header is copied).  \\

\index{PARAMETERS NAME}
PARAMETERS\_NAME & -   & string & The name of the file containing the list of parameters that will be computed and put in the catalog for each object. \\

\hline
\end{tabular*}\\

\subsection{\label{sec:associn}ASSOC parameters}

These are the parameters dealing with crosscorrolating two catalogs: 
a catalog of targets and the output catalog.
(The target catalog is given in ASSOC\_NAME, the output one is created as SE runs).
The cross correlation is controlled by two parameter: ASSOC\_RADIUS 
and ASSOC\_TYPE, the first governing the search radius and latter which 
objects gets selected if there are multiple candidates near the positions.
The numbers of the columns in target catalog which contain the x,y 
positions of the objects need to be in ASSOC\_PARAMS. A third column 
here can be used as weight. So you can crosscorellate catalogs weighted 
with magnitude but also with another parameter such as FWHM or no of 
pixels. Useful it you're crosscorellating between catalogs of different filters.
Some of the columns in target catalog (with the to-be-crosscorellated objects)
can be put into the second output catalog. These columns are specified in 
ASSOC\_DATA and end up in the VECTOR\_ASSOC.

\begin{center}
\noindent\fbox{\parbox{0.8\textwidth}{
NOTE: ASSOC works only with pixel positions (NOT RA and DEC!)
}}
\end{center}

\begin{center}
\noindent\fbox{\parbox{0.8\textwidth}{
NOTE: Be aware of shifts and rotation between images when crosscorrellating two catalogs.
Find shifts beween images with imcentroid in IRAF for instance.}}
\end{center}

\begin{center}
\noindent\fbox{\parbox{0.8\textwidth}{
NOTE: ASSOC will appear not to work if you don't ask for a ASSOC output 
parameter in the parameter file. Your catalog better contain either the 
VECTOR\_ASSOC or the NUMBER\_ASSOC output, otherwise SE simply 
runs and outputs {\it all} objects it has detected, not just the crosscorellated 
ones\footnote{Yes I ended up fiddling with these inputparameters for the 
longest time mumbling "It doesn't WORK!" while I should have asked for 
either VECTOR\_ASSOC or the NUMBER\_ASSOC in the param file. }.
}}
\end{center}

\begin{center}
\noindent\fbox{\parbox{0.8\textwidth}{
Emmanuel sez: Unfortunately, SExtractor can only handle floating point 
numbers in ASSOC files. In fact, ASSOC file may have comment lines 
(like an SE header) but no tab spacing either. Your best bet is just a 
list of x and y positions (with possibly a weight) separated with white 
spaces and nothing else.
}}
\end{center}

\begin{tabular*}{5.0in}[l]{l l p{0.85in} p{1.3in} }
Parameter &	Default &	Type &	Description \\
\hline

\index{ASSOC NAME}
ASSOC\_NAME &  sky.list  & string & Name of the ASSOC ASCII file. \\

\index{ASSOC PARAMS}
ASSOC\_PARAMS &  2,3,4 &  integers ($n \leq 2$, $n \leq 3$) &  Nos of the columns in the ASSOC file that will be used as coordinates and weight for cross-matching. \\

\index{ASSOC RADIUS}
ASSOC\_RADIUS &  2.0 &  float &  Search radius (in pixels) for ASSOC. \\

\index{ASSOC TYPE}
ASSOC\_TYPE  & MAG\_SUM &  keyword  & Method for crossmatching in ASSOC:\\

 & & FIRST  	& keep values corresponding to the first match found, \\
 & & NEAREST 	& values corresponding to the nearest match found, \\
 & & MEAN  	& weighted-average values, \\
 & & MAG\_MEAN  & exponentially weighted average values, \\
 & & SUM  	& sum values, \\
 & & MAG\_SUM  	& exponentially sum values, \\
 & & MIN  	& keep values corresponding to the match with minimum weight, \\
 & & MAX  	& keep values corresponding to the match with maximum weight.\\ 

\hline
\end{tabular*}\\

\begin{tabular*}{5.0in}[l]{l l p{0.8in} p{1.3in} }
Parameter &	Default &	Type &	Description \\
\hline

\index{ASSOCSELEC TYPE}
ASSOCSELEC\_TYPE &  MATCHED  & keyword &  What sources are printed in the out-put catalog in case of ASSOC: \\
 & & ALL  &   all detections,  \\
 & & MATCHED  &   only matched detections, \\
 & &  -MATCHED  &   only detections that were not matched.  \\

\index{ASSOC DATA}
ASSOC\_DATA  & 2,3,4  & integers ($n \leq 32$) &  Numbers of the columns in the ASSOC file that will be copied to the catalog out-put. \\

\hline
\end{tabular*}\\

\begin{center}
\noindent\fbox{\parbox{0.8\textwidth}{
NOTE: ASSOC does not work as long as you do not ask for the ASSOC output in the parameter file (specified in PARAMETERS\_NAME.). So don't forget to ask for VECTOR\_ASSOC or  NUMBER\_ASSOC in the parameter file or otherwise all the detections are reported.
}}
\end{center}

\subsection{\label{sec:check}The Check-images}

\index{CHECKIMAGE}
SE can output some of the maps used in intermediate steps. The names for 
the output fits files are specified in CHECKIMAGE\_NAME. 
Keep a convention like your file\_background.fits or your file\_bgr.fits. 
A list of up to 16 can be given (separated by a comma).\footnote{Oddly enough, this more than covers all your options, including NONE for the type of checkimage.} 

The type of output files you want is defined in CHECKIMAGE\_TYPE.
As you can see, most of these have to do with the background estimation. 
Notable exceptions are the APERTURES and SEGMENTATION options.  
APERTURES is a good diagnostic on whether or not your threshold is 
right and the SEGMENTATION will show you if the objects are broken up 
too much or not. Load the original and this segmentation image into 
saotng and compare.

SEGMENTATION has another useful feature, the number in the catalog is 
given as the value to the isoarea in this image. Good for figuring 
out what is what from the catalogs. Also, the segmentation image is 
used as input for follow-up fit programs of extended sources (GALFIT and GIM2D).

\begin{center}
\noindent\fbox{\parbox{0.8\textwidth}{
NOTE: APERTURES checkimage shows the apertures you've asked for. 
So if you ask for a MAG\_AUTO, the Kron radius will be drawn on the 
checkimage. If NO Kron derived parameter is put into the catalog, the 
Kron radius will also not be drawn on the APERTURES checkimage. 
Same for the APER and PETRO apertures.
}}
\end{center}

\begin{tabular*}{5in}[l]{l l p{0.8in} p{1.4in} }
Parameter &	Default &	Type &	Description \\
\hline

\index{CHECKIMAGE NAME}
CHECKIMAGE\_NAME  & check.fits & strings ($n \leq 16$) & File name for each  check-image . \\
\hline
\end{tabular*}

\begin{tabular*}{5in}[l]{l l p{1.4in} }
Parameter &	Type &	Description \\
\hline
\index{CHECKIMAGE TYPE}
CHECKIMAGE\_TYPE & keywords ($n \leq 16$) & Type of information to put in the  check-images : \\
 &  NONE &  no check-image, \\
 &  IDENTICAL & identical to input image (useful for converting formats),\\ 
 &  BACKGROUND & full-resolution interpolated background map, \\
 &  BACKGROUND\_RMS & full-resolution interpolated background noise map, \\
 &  MINIBACKGROUND & low-resolution background map, \\
 &  MINIBACK\_RMS & low-resolution background noise map, \\
 &  -BACKGROUND & background-subtracted image, \\
 &  FILTERED &  background-subtracted filtered image (requires FILTER = Y),\\
 &  OBJECTS &  detected objects, \\
 &  -OBJECTS &  background-subtracted image with detected objects blanked, \\
 &  APERTURES &  MAG APER and MAG AUTO integration limits, \\
 &  SEGMENTATION & display patches corresponding to pixels attributed to each object. \\
\hline
\end{tabular*}\\

\begin{figure}[htb]
    \centering
    \caption{Illustration of the different checkimages possible. The original inputimage, the BACKGROUND image, the FILTERED image, the OBJECTS image, the SEGMENTATION image and the APERTURES image. the contrast of the BACKGROUND image has been exaggerated. }
    \label{fig:check}
\end{figure}

\chapter{\label{ch:output}Output Parameters}

The catalogs with output parameters is what the whole exercise is all about!\footnote{Well that and the check-images perhaps.}
You can finally start constructing your Hertzsprung-Russel diagrams or 
lensing shear fields or whatever. The parameters you want in your catalogs 
should be listed in the file you gave to PARAMETERS\_NAME. Unless you keep 
using the same file for this, I really recommend using ASCII\_HEAD type 
catalogs \footnote{By setting the parameter CATALOG\_TYPE to ASCII\_HEAD 
in the configuration file.} The output catalogs will have a nice header with a list of all
the parameters. \\

Parameters in the SE outputcatalog can be divided into geometric 
parameters ,photometric parameters, astrometric parameters and fitted 
parameters.\footnote{Just kidding, SE does not fit anything yet. It's so fast for a reason.} 
Geometric parameters will tell you what shape the object 
is in (basically how the light of the object is distributed over the 
pixels of that object) and the photometric parameters tell you simply 
how much light there is. Astrometric parameters give the position of 
the object in the image, be it in pixels or other coordinates.\footnote{There has been some improvement here as well. The now nearly ubiquitous World Coordinate System WCS is also used.}
Fitted parameters are calculated from fitting for instance a PSF to 
the data of the object. Most of these are still being developed and do not work yet.\footnote{Meaning, I've tried but have not gotten them to work yet...If you have, let me know.}

There are a few that do not fall in any of these convenient categories; the catalog number, the flag parameters and the parameters associated with crosscorrellating catalogs.

\begin{tabular*}{5in}[l]{l l l }

Name 		& description 					& unit \\
\hline

\index{NUMBER}
 NUMBER		& Running object number  			& - \\

\index{FLAGS}
 FLAGS		& Extraction flags					& - \\
\index{IMAFLAGS ISO}
 IMAFLAGS\_ISO	& FLAG-image flags OR'ed over the iso. profile	  	& - \\
\index{NIMAFLAGS ISO}
 NIMAFLAGS\_ISO	& \# flagged pixels entering IMAFLAGS\_ISO  	& - \\
\hline
\end{tabular*}\\

\section{\label{sec:photout}Photometric Parameters}

Photometric parameters are either flux or magnitude determined by SE.
However SE has five different ways of determining these; isophotal, 
isophotal-corrected, automatic, best estimate and aperture.
These are discussed in the Photometry section in the input chapter (section \ref{sec:photin}).
To recap:\\

\begin{tabular*}{5in}[l]{l p{3.5in}}

ISO & Photometry derived from the counts above the threshold minus the background (see also section \ref{sec:iso}). \\

ISOCOR & ISO photometry, corrected for loss as a Gaussian profile (see also section \ref{sec:isocor}).\\

AUTO & Photometry from the Kron flexible elliptical aperture. \cite{Kron80} (see also section \ref{sec:auto})\\

BEST & Choice between AUTO and ISOCOR. AUTO, except when influence from neighbors is more than 10\%. (see also section \ref{sec:best}) \\

APER & Photometry from circular, user specified (PHOT\_APERTURES in the config file), apertures. (see also section \ref{sec:aper})\\

PETRO & Photometry from the Petrosian aperture, very similar to the Kron aperture. (see also section \ref{sec:})\\
PROFILE & the weighted photometry using the 'filtered' image for the weight.
\end{tabular*}\\

\begin{figure}[htb]
    \centering
    \caption{Illustration of the different apertures possible; ISO, ISOCOR, AUTO and APER (user specified in PHOT\_APERTURES)}
    \label{fig:aper2}
\end{figure}

There are two other photometric parameters of interest: MU\_MAX, the surface
brightness of the brightest pixel and the MU\_THRESHOLD, the the surface
brightness corresponding to the threshold. This last parameter is good to 
inspect if the threshold is set with respect to the background RMS.
The value of the Background at the position of the object is often also interesting to know. Especially when deciding whether or not to switch between GLOBAL and LOCAL in the BACK\_TYPE parameter (see section \ref{sec:back}).

The Petrosian aperture is a recent addition to SE (since v2.4.4 as far as I know) and it is very similat to the Kron radius. The apreture has a different radius (a Petrosain vs. a Kron one.). However the position angle and the ellipticity are the same as the Kron aperture. The Petrosian radius is usually bigger than the Kron one. See also section \ref{sec:inR1}, \ref{sec:R1} and \ref{sec:auto} for Kron aperture stuff and see section \ref{sec:inRp} and \ref{sec:Rp} for more on the Petrosian radius.

\begin{center}
\noindent\fbox{\parbox{0.8\textwidth}{
NOTE: oh and I'll say it again...the "BEST" photometry is a misnomer. Rather use something that is consistent across your image.
}}
\end{center}

\begin{center}
\noindent\fbox{\parbox{0.8\textwidth}{
NOTE: For a color (Great Brittain and Canada: colour) measurement, the ISO and APER options are good, especially when run in dual mode since you'll know the apertures are the same. The other apertures may be too inclusive in crowded fields.
}}
\end{center}


\begin{tabular*}{5in}[l]{l p{2.7in} p{0.5in} }

Name 		& description 					& unit \\
\hline
\index{FLUX ISO}
 FLUX\_ISO	& Isophotal flux   				& count   \\
\index{FLUXERR ISO}
 FLUXERR\_ISO	& RMS error for isophotal flux			& count   \\
\index{MAG ISO}
 MAG\_ISO	& Isophotal magnitude   			& mag   \\
\index{MAGERR ISO}
 MAGERR\_ISO	& RMS error for isophotal magnitude		& mag   \\
& & \\

\index{FLUX ISOCOR}
 FLUX\_ISOCOR	& Corrected isophotal flux   			& count  \\ 
\index{FLUXERR ISOCOR}
 FLUXERR\_ISOCOR	& RMS error for corrected isophotal flux	& count  \\ 
\index{MAG ISOCOR}
 MAG\_ISOCOR	& Corrected isophotal magnitude			& mag   \\
\index{MAGERR ISOCOR}
 MAGERR\_ISOCOR	& RMS error for corrected isophotal magnitude	& mag  \\ 
& & \\

\index{FLUX AUTO}
 FLUX\_AUTO	& Flux within a Kron-like elliptical aperture	& count  \\ 
\index{FLUXERR AUTO}
 FLUXERR\_AUTO	& RMS error for AUTO flux			& count  \\ 
\index{MAG AUTO}
 MAG\_AUTO	& Kron-like elliptical aperture magnitude	& mag   \\
\index{MAGERR AUTO}
 MAGERR\_AUTO	& RMS error for AUTO magnitude			& mag   \\
& & \\

\index{FLUX BEST}
 FLUX\_BEST	& Best of FLUX\_AUTO and FLUX\_ISOCOR   		& count  \\ 
\index{FLUXERR BEST}
 FLUXERR\_BEST	& RMS error for BEST flux			& count  \\ 
\index{MAG BEST}
 MAG\_BEST	& Best of MAG\_AUTO and MAG\_ISOCOR		& mag  \\ 
\index{MAGERR BEST}
 MAGERR\_BEST	& RMS error for MAG\_BEST			& mag   \\
& & \\

\index{FLUX APER}
 FLUX\_APER	& Flux vector within fixed circular aperture(s)	& count \\
\index{FLUXERR APER}
 FLUXERR\_APER	& RMS error vector for aperture flux(es)   	& count  \\
\index{MAG APER}
 MAG\_APER	& Fixed aperture magnitude vector		& mag \\
\index{MAGERR APER}
 MAGERR\_APER	& RMS error vector for fixed aperture mag.	& mag \\
& & \\

\index{FLUX PETRO}
 FLUX\_PETRO	& Flux within a Petrosian-like elliptical aperture & count \\
\index{FLUXERR PETRO}
 FLUXERR\_PETRO	& RMS error for PETROsian flux    	& count  \\
\index{MAG PETRO}
 MAG\_PETRO	& Petrosian-like elliptical aperture magnitude	& mag \\
\index{MAGERR PETRO}
 MAGERR\_PETRO	& RMS error for PETROsian magnitude  	& mag \\
& & \\

\hline
\end{tabular*}\\

\section{\label{sec:profile}Profile}

The profile option for photometry is not a much used one. The idea is to 
weigh the fluxes with the values in the filtered (smoothed) image (see 
section \ref{sec:filter} for the options here). in general, smoothing correlates 
the noise in an image and broadens profiles. By weighting with the smoothed 
profile, you give extra weight to the brightest pixels (as opposed to counting 
all the flux equally). 

\begin{center}
\noindent\fbox{\parbox{0.8\textwidth}{
NOTE: the PROFILE option seems to be a fix for when you think you include 
too much noise in the flux measurements (i.e. the detection threshold is too 
low, background too funky or something...)
Use with some caution. 
}}
\end{center}

\begin{center}
\noindent\fbox{\parbox{0.8\textwidth}{
And remember kids: the ellipticity and everything is derived from the 'filtered' 
(i.e. smoothed) image. The PROFILE option in this context might be a good 
way to check up how much different things are between the smoothed and 
the original image.
}}
\end{center}

\begin{tabular*}{5in}[l]{l p{2.7in} p{0.5in} }

Name 		& description 					& unit \\
\hline
\index{FLUX PROFILE}
 FLUX\_PROFILE	& Flux weighted by the FILTERed profile		& count \\  
\index{FLUXERR PROFILE}
 FLUXERR\_PROFILE &  RMS error for PROFILE flux			& count \\  
\index{MAG PROFILE}
 MAG\_PROFILE	& Magnitude weighted by the FILTERed profile	& mag   \\
\index{MAGERR PROFILE}
 MAGERR\_PROFILE	& RMS error for MAG\_PROFILE			& mag   \\
\hline
\end{tabular*}\\

\begin{tabular*}{5in}[l]{l p{2.2in} p{1.2in} }
Name 		& description 					& unit \\
\hline
\index{MU THRESHOLD}
 MU\_THRESHOLD	& Detection threshold above background			& mag $\times$ arcsec${-2}$ \\   
\index{MU MAX}
 MU\_MAX		& Peak surface brightness above background	& mag $\times$ arcsec$^{-2}$ \\   

\index{BACKGROUND}
BACKGROUND & Background at centroid position & counts \\
\index{THRESHOLD}
THRESHOLD  & Detection threshold above background & counts\\

\hline
\end{tabular*}\\

\section{\label{sec:xy}Astrometric Parameters}

The astrometric parameters are simple enough; they tell you where the object
 is located. However which pixel do you take for the center of the objects? 
The maximum flux pixel? Or the barycenter?
\footnote{The Barycenter is the flux-weighted average position. the first moment, SE computes all the second moments as well, see the geometric parameters section and chapter 9 in the manual v2.1.3.}
 Do you want it in RA and DEC or in prosaic pixels? All of these you should 
be able to specify from the list below.\\

\begin{center}
\noindent\fbox{\parbox{0.8\textwidth}{
NOTE: the x and y positions (with the exception of 'peak' values) are the barycenters of objects; the weighted mean position.
}}
\end{center}

\index{barycenter}
The definition for the barycenter of an object is:

$$ X = \overline{x} = {\Sigma I_i x_i \over \Sigma I_i}$$
$$ Y = \overline{y} = {\Sigma I_i y_i \over \Sigma I_i}$$

\noindent This is the first order moment of the object.\footnote{Yes this is straight 
from the sextractor manual. All this stuff is quite complete in the manual 
v2.1.3 and above. Is there a problem?}
The minima and maximum x and y pixelvalues almost speak for themselves.
These are the most extreme values for x and y that are still within the object. 
Handy if you want to make postage stamps of some of your objects from the fits file.
The PEAK values are the x,y and ra and dec values for the position of the 
brightest pixel. The value for the flux and surface brightness of that pixel 
can be found in output-parameters FLUX\_MAX and MU\_MAX respectively. 
\index{FLUX MAX}\index{MU MAX}
The difference between the position of the brightest pixel and the (bary) center 
of  an object might help you identify blended objects. If they differ much, than 
you've got something lobsided. 

\begin{center}
\noindent\fbox{\parbox{0.8\textwidth}{
NOTE: all the $x_i$ and $I_i$ values in these formulae are the values from the pixels identified in the segmentation map as belonging to the object. It can be influenced by detection thresholds and segmentation settings.
}}
\end{center}

\begin{center}
\noindent\fbox{\parbox{0.8\textwidth}{
NOTE: The origin for X\_IMAGE etc is pixel 1,1. SO The first pixel in the left bottom corner of the FITS image is number 1,1 and not 0,0.
}}
\end{center}

\begin{tabular*}{5.2in}[l]{l p{2.7in} l }

Name & description & unit \\
\hline
\index{XMIN IMAGE}
 XMIN\_IMAGE		& Minimum x-coordinate among detected pixels	& pixel \\  
\index{YMIN IMAGE}
 YMIN\_IMAGE		& Minimum y-coordinate among detected pixels	& pixel \\  
\index{XMAX IMAGE}
 XMAX\_IMAGE		& Maximum x-coordinate among detected pixels	& pixel \\  
\index{YMAX IMAGE}
 YMAX\_IMAGE		& Maximum y-coordinate among detected pixels	& pixel\\   

\index{XPEAK IMAGE}
 YPEAK\_IMAGE		& y-coordinate of the brightest pixel		& pixel\\   
\index{XPEAK WORLD}
 XPEAK\_WORLD		& World-x coordinate of the brightest pixel	& deg   \\
\index{YPEAK WORLD}
 YPEAK\_WORLD		& World-y coordinate of the brightest pixel	& deg   \\

\index{ALPHAPEAK SKY}
 ALPHAPEAK\_SKY		& Right ascension of brightest pix (native)	& deg  \\ 
\index{DELTAPEAK SKY}
 DELTAPEAK\_SKY		& Declination of brightest pix (native)		& deg   \\

\index{ALPHAPEAK J2000}
 ALPHAPEAK\_J2000	& Right ascension of brightest pix (J2000) 	& deg \\  
\index{DELTAPEAK J2000}
 DELTAPEAK\_J2000	& Declination of brightest pix (J2000)	 	& deg  \\ 

\index{ALPHAPEAK B1950}
 ALPHAPEAK\_B1950	& Right ascension of brightest pix (B1950) 	& deg \\  
\index{DELTAPEAK B1950}
 DELTAPEAK\_B1950	& Declination of brightest pix (B1950)	 	& deg   \\

\index{X IMAGE}
 X\_IMAGE		& Object position along x			& pixel \\   
\index{Y IMAGE}
 Y\_IMAGE		& Object position along y			& pixel \\   
\index{X IMAGE DBL}
 X\_IMAGE\_DBL		& Object position along x (double precision)	& pixel  \\ 
\index{Y IMAGE DBL}
 Y\_IMAGE\_DBL		& Object position along y (double precision)	& pixel  \\ 
\index{X WORLD}
 X\_WORLD		& Barycenter position along world x axis	& deg \\  
\index{Y WORLD}
 Y\_WORLD		& Barycenter position along world y axis	& deg \\  
\index{X MAMA}
 X\_MAMA		& Barycenter position along MAMA x axis		& m$^{-6}$  \\ 
\index{Y MAMA}
 Y\_MAMA		& Barycenter position along MAMA y axis		& m$^{-6}$  \\

\index{ALPHA SKY}
 ALPHA\_SKY		& Right ascension of barycenter (native)	& deg   \\
\index{DELTA SKY}
 DELTA\_SKY		& Declination of barycenter (native)		& deg  \\ 

\index{ALPHA J2000}
 ALPHA\_J2000		& Right ascension of barycenter (J2000)		& deg \\  
\index{DELTA J2000}
 DELTA\_J2000		& Declination of barycenter (J2000)		& deg \\  

\index{ALPHA B1950}
 ALPHA\_B1950		& Right ascension of barycenter (B1950)		& deg \\  
\index{DELTA B1950}
 DELTA\_B1950		& Declination of barycenter (B1950)		& deg \\  
\hline
\end{tabular*}\\

According to the v2.3 manual of SE, FITS header-information is used for the WORLD coordinates.
World Coordinates System (WCS) is often what you use to line up different images from different observatories in ds9 (that's what I do...). \\
No idea what the MAMA axis is.

\section{\label{sec:geom}Geometric Parameters}

Geometric parameters describe the shape and size of the object. 
SE computes the moments of an object and determines elliptical 
parameters from these. Both the moments and the elliptical parameters 
and derivatives from these can be included in the output.
These parameters are treated much more extensive in chapter 9 of the manual 
(version 2.1.3).

\subsection{\label{sec:moments}Moments}

The first order moments are the barycenters of course but the second order 
moments can also be given (see section 9.1.4 in the manual v2.1.3):

\begin{equation}
X2 = \overline{x^2} = {\Sigma I_i x_i^2 \over \Sigma I_i} - \overline{x}^2
\end{equation}

\begin{equation}
Y2 = \overline{y^2} = {\Sigma I_i y_i^2 \over \Sigma I_i} - \overline{y}^2
\end{equation}

\begin{equation}
XY = \overline{xy} = {\Sigma I_i x_i y_i \over \Sigma I_i} - \overline{xy}
\end{equation}

\begin{tabular*}{5in}[l]{l l l }
Name & description & unit \\
\hline
\index{X2 IMAGE}
  X2\_IMAGE	& Variance along x				& pixel$^2$   \\
\index{Y2 IMAGE}
  Y2\_IMAGE	& Variance of position along y				& pixel$^2$   \\ 
\index{XY IMAGE}
  XY\_IMAGE	& Covariance of position between x and y		& pixel$^2$  \\  
\index{X2 WORLD}
  X2\_WORLD	& Variance of position along X-WORLD (alpha)		& deg$^2$ \\   
\index{Y2 WORLD}
  Y2\_WORLD	& Variance of position along Y-WORLD (delta)		& deg$^2$  \\  
\index{XY WORLD}
  XY\_WORLD	& Covariance of position X-WORLD/Y-WORLD		& deg$^2$ \\   
\hline
\end{tabular*}\\

\begin{center}
\noindent\fbox{\parbox{0.8\textwidth}{
NOTE: By adding ERR+parameter, the error can be obtained as well. (e.g. ERRX2\_IMAGE)
}}
\end{center}

\subsection{\label{sec:ellipse}Ellipse parameters}

From these moments, the position angle, the minor and major axis of an ellipse can be
derives (section 9.1.5 in the manual v2.3) and also a second way of describing this 
ellipse is given (sec 9.1.6). The official manual gives the whole derivation of these 
parameters but since we want a quick result I'll cut it down some.

\begin{center}
\noindent\fbox{\parbox{0.8\textwidth}{
NOTE: All these elliptical parameters are computed in the SMOOTHED image. So If you 
have used a smoothing kernel in the input, then the ELLIPTICITY is also of a 
smoothed object. \footnote{This little fun fact had an astronomer of note almost smack his head out frustration against a bookshelf in the STSCI library. It explained why he was seeing not enough high-ellipticity galaxies in a deep field, they were rounded by the smoothing. Do not let this happen to you. Be wary of bookshelves...}
}}
\end{center}

\subsubsection{\label{sec:axes}Minor and Major Axes}

The following definitions for these {\bf computed} parameters are:
(see for the derivation the v2.3 manual page 28)

\begin{equation}
A^2 = { \overline{x^2} + \overline{x^2} \over 2} + \sqrt{ \left({ \overline{x^2} - \overline{y^2} \over 2} \right)^2 +\overline{xy}^2} 
\end{equation}

\begin{equation}
B^2 = { \overline{x^2} + \overline{x^2} \over 2} - \sqrt{ \left({ \overline{x^2} - \overline{y^2} \over 2} \right)^2 +\overline{xy}^2}
\end{equation}

These are illustrated in figure \ref{fig:kron}. Not also that the Petrosian and Kron radii (section \ref{sec:Rp} and \ref{sec:R1}) are expressed in the same units as those used for the A and B values.

\begin{figure}[htb]
    \centering
    \caption{Illustration of the Kron radius. $\bar{x}$ and $\bar{y}$ are the computed center in of the object. }
    \label{fig:kron}
\end{figure}

\begin{tabular*}{5in}[l]{l l l }
Name & description & unit \\
\hline
\index{A IMAGE}
 A\_IMAGE	& Profile RMS along major axis				& pixel \\   
\index{B IMAGE}
 B\_IMAGE	& Profile RMS along minor axis				& pixel \\   
\index{THETA IMAGE}
 THETA\_IMAGE	& Position angle (CCW/x) {\bf counterclockwise!!!}\footnote{I've noticed that the image in the official manual now also reflects this...so I copied it in figure \ref{fig:kron}.}
 	& deg  \\  
\index{A WORLD}
 A\_WORLD	& Profile RMS along major axis (world units)		& deg  \\  
\index{B WORLD}
 B\_WORLD	& Profile RMS along minor axis (world units)		& deg  \\  
 & & \\
 \hline
\end{tabular*}\\
\noindent Errors can be obtained using ERR+parameter in the paramfile.

\subsubsection{\label{sec:pa}Position Angle}

This is the rotation of the major axis with respect to NAXIS1 (x-axis...) counterclockwise.
The definition of the position angle is different (pa is with respect to North.)
However the conversion can obviously be done and the Position Angle (PA) is given in other THETA parameters (SKY, B1950 and J2000). This can be very useful if you're looking for slit-positions for your spectrograph.

\begin{tabular*}{5in}[l]{l l l }
Name & description & unit \\
\hline 
\index{THETA WORLD}
 THETA\_WORLD	& Position angle (CCW/world-x)				& deg  \\  
\index{THETA SKY}
 THETA\_SKY	& Position angle (east of north) (native)		& deg  \\  
\index{THETA J2000}
 THETA\_J2000	& Position angle (east of north) (J2000)		& deg   \\ 
\index{THETA B1950}
 THETA\_B1950	& Position angle (east of north) (B1950)		& deg   \\ 
 & & \\
\hline
\end{tabular*}

\noindent Errors can be obtained using ERR+parameter in the paramfile.

\subsubsection{\label{sec:pa}The Other Ellipse parametrisation}

So the manual also gives another parametrisation of the elliptical aperture. I have not used these parameters. I have no idea what to do with them. Maybe you do. In figure \ref{fig:kron}, the relation between these parameters and the moments is given.

\begin{tabular*}{5in}[l]{l l l }
Name & description & unit \\
\hline 
\index{CXX}
  CXX\_IMAGE	& Cxx object ellipse parameter			& pixel$^{-2}$   \\
\index{CYY}
  CYY\_IMAGE	& Cyy object ellipse parameter			& pixel$^{-2}$   \\
\index{CXY}
  CXY\_IMAGE	& Cxy object ellipse parameter			& pixel$^{-2}$   \\
  CXX\_WORLD	& Cxx object ellipse parameter (WORLD units)	& deg$^{-2}$   \\
  CYY\_WORLD	& Cyy object ellipse parameter (WORLD units)	& deg$^{-2}$   \\
  CXY\_WORLD	& Cxy object ellipse parameter (WORLD units)	& deg$^{-2}$   \\
 & & \\
 \hline
\end{tabular*}

\noindent Errors can be obtained using ERR+parameter in the paramfile.

\subsubsection{\label{sec:ell}Ellipticity and Elongation}
 
How stretched the object is can be parameterised with weither the Ellipticity 
or the Elongation parameters, both of which are simply computed from the 
minor and major axes.
 
\begin{equation}
\rm ELONGATION = {A \over B}
\end{equation}

\begin{equation}
\rm ELLIPTICITY =  1 - {B \over A} = 1 - {1 \over ELONGATION}
\end{equation}

\begin{tabular*}{5in}[l]{l l l }
Name & description & unit \\
\hline 
\index{ELONGATION}
 ELONGATION	& A\_IMAGE/B\_IMAGE  					& \\
\index{ELLIPTICITY}
 ELLIPTICITY	& 1 - B\_IMAGE/A\_IMAGE  				& \\
\hline
\end{tabular*}\\

\begin{center}
\noindent\fbox{\parbox{0.8\textwidth}{
NOTE: If you have FILTERED the image for detection then your ellipticity and elongation values will be more {\bf round} than the actual object. }}
\end{center}

Only the ELONGATION and the ELLIPTICITY parameters do not come with a error.
Note that there is an extensive discussion of these in section 9.1.8 of the manual.
\footnote{I'd be saying the exact same thing and typing in that many functions is a pain, even in LaTeX.}

\subsection{\label{sec:area}Area Parameters}

This can be done by the parameters describing the isophotes as 
fitted by SE or the conclusion of the neural network classification.
 To get an idea of the size of an object, reasonably independent of 
the brightness, the FWHM or the FLUX\_RADIUS

SE divides an object up into 7 isophotes above the ANALYSIS\_THRESH. 
The areas above the isophotes is fed to the neural network. These can be 
put into the catalog with the ISO{\it n} parameters.

\begin{tabular*}{5in}[l]{l p{2.7in} l }
Name & description & unit \\
\hline

\index{ISOAREA WORLD}
 ISOAREA\_WORLD	& Isophotal area above Analysis threshold		& deg$^2$    \\
\index{ISOAREAF WORLD}
 ISOAREAF\_WORLD	& Isophotal area (filtered) above Detection threshold 	& deg$^2$    \\
\index{ISO0}
 ISO0	& Isophotal area at level 0					& pixel$^2$    \\
\index{ISO1}
 ISO1	& Isophotal area at level 1					& pixel$^2$    \\
\index{ISO2}
 ISO2	& Isophotal area at level 2					& pixel$^2$    \\
\index{ISO3}
 ISO3	& Isophotal area at level 3					& pixel$^2$   \\ 
\index{ISO4}
 ISO4	& Isophotal area at level 4					& pixel$^2$   \\ 
\index{ISO5}
 ISO5	& Isophotal area at level 5					& pixel$^2$   \\ 
\index{ISO6}
 ISO6	& Isophotal area at level 6					& pixel$^2$    \\
\index{ISO7}
 ISO7	& Isophotal area at level 7					& pixel$^2$    \\

\hline
\end{tabular*}\\

\subsection{\label{sec:fwhm}Full-Width Half Max}

Assuming a Gaussian profile for the object, a Full-Width at Half Maximum can be computed. 
No object is a Gaussian -especially in the wings- but this does give you a reasonable idea of the PSF width in the case of a star for instance. It's a useful thing to have.

\begin{tabular*}{5in}[l]{l p{2.7in} l }
Name & description & unit \\
\hline
\index{FWHM IMAGE}
 FWHM\_WORLD	& FWHM assuming a Gaussian core				& pixel   \\
\index{FWHM WORLD}
 FWHM\_WORLD	& FWHM assuming a Gaussian core				& deg   \\

\index{VIGNET}
 VIGNET		& Pixel data around detection				& count \\
\index{VIGNET SHIFT}
 VIGNET\_SHIFT	& Pixel data around detection corrected for shift	& count \\

\index{THRESHOLDMAX}
 THRESHOLDMAX	& Maximum threshold possible for detection		& count  \\ 

\hline
\end{tabular*}\\

\section{\label{sec:radii} Radii}

Indicators for the size of objects are multiple and SE can provide a lot of them if so desired. 
Especially for extended source several schemes have been developed to define an aperture to determine the total flux and characterize the size of an object. In the section \ref{sec:inR1}, \ref{sec:ellipse} and \ref{sec:auto}, the Kron aperture and the various associated parameters were already discussed. The KRON\_RADIUS output parameter is the indicator of size of the KRON aperture. Alternatively, there is the PETROSIAN\_RADIUS (since SE v2.4.4).  Both of these are expressed in multiples of major axis B.

\begin{table}[htdp]
\caption{Radii nomenclature}
\begin{center}
\begin{tabular}{l l l}
Radius	& symbol	& Description \\
\hline
\hline
Kron 	& $R_1$	& The typical size of the flexible aperture -computed from moments- \\
		&		& defined by \cite{Kron80}.\\
Petrosain	& $R_p$	& The radius at which the surface brightness of the isophote is \\
		&		& $\eta$ times the average surfacebrightness within this isophote.\\
Effective	& $R_e$	& The radius containing 50\% of the total flux of an object.\\
Half-light	&		& same radius.\\
Typical	& $R_t$ or h & The scale in the exponential disk $I ~ = ~ I_0 ~ exp(-R/R_t)$\\
deVaucouleur & $R_{25}$ & The radius at which the B-band isophote is \\
		&		& $\rm 25 ~ mag ~ arcsec^{-2}$. This is a common Radius as \\
		&		& it is listed in the third reference catalog (RC3)\\
\hline
\end{tabular}
\end{center}
\label{table:radii}
\end{table}%

\subsection{\label{sec:R1}Kron radius}

\cite{Kron80} defined the KRON radius to get 90\% of an objects light as follows:

\begin{equation}
R_1(R) = {2 \pi \int_0^R I(x) x^2 dx \over 2 \pi \int_0^R I(x) x dx}
\end{equation}

According \cite{Graham05a}, the orginal SE publication has changed the definition a bit.
\cite{SE} define the Kron radius as:

\begin{equation}
R_1 = { \Sigma R ~I(R) \over \Sigma I(R)} 
\end{equation}

\noindent over the two dimentional aperture and not for a lightprofile (which was Kron's original idea.)
In the case of a lightprofile, the Kron-radius should be according to Kron's original definition:

\begin{equation}
R_1 = { \Sigma R^2 ~ I(R) \over \Sigma R ~ I(R)} 
\end{equation}

So the Kron radius coming out of SE is maybe not a proper one but we're 
using this as the estimate anyway. And if you are interested in galaxy 
profiles, here is a neat trick to relate this Kron radius to 
the {\it effective} radius ($R_e$), the radius that encloses half the object's 
light.\footnote{All these radii to characterize galaxy profiles can be a tad 
confusing. See table \ref{table:radii} for the definitions of all.}

\begin{equation}
R_1 (x,n) = {R_e \over b^n} {\gamma(3n,x) \over \gamma(2n,x)}
\end{equation}

\noindent In which $x=b(R/R_e)^{1/n}$ and $\gamma$ is the incomplete 
gamma function\footnote{Oh Gamma function, is there something you're 
not useful for?}. This is all too much of a headache so \cite{Graham05a} 
also give a nice table converting these radii depending on Sersic profile 
index (copied in table \ref{table:kronradii}). 

\begin{table}[htdp]
\caption{Theoretical Kron Radii and Magnitudes from \cite{Graham05a}}
\begin{center}
\begin{tabular}{l l l l}
Sersic n 	& $R_1$	& $L( <  2R_1)$ & $L(< 2.5R_1)$\\
			& ($R_e$) & \%		& \% \\ 
\hline
0.5 	& 1.06 	& 95.7 	& 99.3\\
1.0 	& 1.19 	& 90.8 	& 96.0\\
2.0 	& 1.48 	& 87.5 	& 92.2\\
3.0 	& 1.84 	& 86.9 	& 90.8\\
4.0 	& 2.29 	& 87.0 	& 90.4\\
5.0 	& 2.84 	& 87.5 	& 90.5\\
6.0 	& 3.53 	& 88.1 	& 90.7\\
7.0 	& 4.38 	& 88.7 	& 91.0\\
8.0 	& 5.44 	& 89.3 	& 91.4\\
9.0 	& 6.76 	& 90.0 	& 91.9\\
10.0 	& 8.39 	& 90.6 	& 92.3\\
\hline
\end{tabular}
\end{center}
\label{table:kronradii}
\end{table}%

\begin{tabular*}{5in}[l]{l p{2.7in} l }
Name & description & unit \\
\hline
\index{KRON RADIUS}
KRON\_RADIUS    & Kron radius in units of A or B & -\\
\hline
\end{tabular*}

\noindent And remember that the KRON flux measurements can be found 
in the AUTO output-parameters.\footnote{Named differently for your inconvenience. 
Please remember that a lot of SE stuff got tacked on later which explains 
the funky naming 'convention'.}

\subsection{\label{sec:Rp}Petrosian radius}

\cite{Petrosian} defined a point in the radial light profile at which the isophote 
at that radius was a certain fraction of the average surface brightness within 
that radius. 

\begin{equation}
\eta(R) = {2 \pi \int_0^R I(R') R' dR' \over \pi R^2 I(R)}  =  {\langle I \rangle_R \over I(R)} 
\end{equation}

\noindent where the parameter $\eta$ is the fraction. This is often either 0.2 or 
0.5 with 0.2 the most commonly used. I am pretty sure the 0.2 value is used in SE.

\begin{center}
\noindent\fbox{\parbox{0.8\textwidth}{
NOTE: The $\eta$ cannot be changed using some input-parameter. Bummer.
}}
\end{center}

There is an input parameter associated with this in source extractor ( 
PHOT\_PETROPARAMS) where the first input defines the number of Petrosian 
radii for the magnitude and the second the minimum radius.
The following output parameters are associated with the Petrosian radius:

\begin{tabular*}{5in}[l]{l p{2.7in} l }
Name & description & unit \\
\hline
\index{PETRO RADIUS}
PETRO\_RADIUS    & Petrosian apertures in units of A or B & -\\
\index{FLUX PETRO}
 FLUX\_PETRO	& Flux within Petrosian radius elliptical aperture	& count  \\ 
\index{FLUX PETRO}
 FLUX\_PETRO	& Flux within Petrosian radius elliptical aperture	& count  \\ 
\index{FLUXERR PETRO}
 FLUXERR\_PETRO	& RMS error for PETRO flux			& count  \\ 
\index{MAG PETRO}
 MAG\_PETRO	& magnitude within the $N \times R_p$	& mag   \\
\index{MAGERR PETRO}
 MAGERR\_PETRO	& RMS error for the MAG\_PETRO magnitude			& mag   \\
& & \\
\hline
\end{tabular*}\\

\noindent More on the Petrosian radius can be found in \cite{Graham05a} and 
\cite{Graham05b}.


\subsection{\label{sec:Re}Effective radius}

The effective radius is the term commonly used to define the point in a light 
profile within which encloses half the flux from an object. In the case of SE, 
several radii can be defined (input parameter PHOT\_FLUXFRAC).
PHOT\_FLUXFRAC 0.5 is the effective radius (or SE's determination of it.)

\begin{tabular*}{5in}[l]{l p{2.7in} l }
Name & description & unit \\
\hline
\index{FLUX RADIUS}
FLUX\_RADIUS & Radius enclosing a specified fraction of the flux & pixel. \\
\hline
\end{tabular*}\\

\noindent With the photometry in the APER parameters. 

So here is another trick. If you's want an estimate of the  Sersic profile 
index n, you could use the ratio between the FLUX\_RADIUS with PHOT\_FLUXFRAC 
set to 0.5 and the KRON\_RADIUS in your catalog.
Of course you would need to run some follow-up program to properly determine 
the actual index from a fit, but that is what SE is all about.\footnote{A quick and possibly dirty first look, quick sample selection etc...}

\section{Object classification}

\index{classification}
This is a section completely devoted to the CLASS\_STAR parameter; SE's 
classification of the objects on the basis of a Neural Network Output. 
\footnote{If you DO want to know more on Neural Networks and how they are used for complex issues such as object classifiacation, I found the book 'An introduction to Neural Networks' by Kevin Gurney very useful.}\\

\begin{tabular*}{5in}[l]{l l l }
Name & description & unit \\
\hline
\index{CLASS STAR}
 CLASS\_STAR	& S/G classifier output  				& none\\
\hline
\end{tabular*}\\

\noindent It can have a value between 0 (galaxy, more to the point: non-star) and 1 (star).

\subsection{Input Dependency}

\index{classification, input}
Fortunately, you do not need to understand Neural Networks to use this but 
there are several input parameters which are directly linked to the 
CLASS\_STAR parameter: 

\begin{itemize}
\index{FLAG IMAGE}
\item PIXEL\_SCALE: Pixel size in arcsec. (for surface brightness parameters, FWHM and star/ galaxy separation only).
\index{SEEING FWHM}
\item SEEING\_FWHM: FWHM of stellar images in arcsec. This quantity is used only for the neural network star/galaxy separation as expressed in the\\ CLASS\_STAR output.
\end{itemize}

These are obvious. The NNW has to take in account the scale and blurring of the objects before judgment. These are inputparameters of the NNW. The ratio between the two is something you can play around with. (remember that SE does not use the pixelsscale for anything else...)
 But there are parameters where it depend more indirectly on:

\begin{itemize}
\index{BACK SIZE}
\item BACK\_SIZE Size, or Width, Height (in pixels) of a background mesh.
\index{THRESH TYPE}
\item THRESH\_TYPE Meaning of the DETECT THRESH and ANALYSIS\_THRESH parameters : 
\begin{itemize}
\item RELATIVE scaling factor to the background RMS.
\item ABSOLUTE absolute level (in ADUs or in surface brightness).
\end{itemize}
\index{ANALYSIS THRESH}
\item ANALYSIS\_THRESH Threshold (in surface brightness) at which CLASS STAR and FWHM operate.
\begin{itemize}
\item 1 argument: relative to Background RMS. 
\item 2 arguments: mu ($\rm mag arcsec^{-2}$ ), Zero-point (mag).\\
\end{itemize}
\end{itemize}

Obviously the brightness level from which objects are considered influence 
heavily the classification. If only the top levels are considered, even 
clear galaxies might be classified as stars (the disk lies below the 
threshold) and if the threshold is too low, random noise can pass itself 
off as a faint galaxy.

\begin{center}
\noindent\fbox{\parbox{0.8\textwidth}{
NOTE: if the threshold is relative to the RMS, the 
BACK\_SIZE is VERY important for the CLASS\_STAR
}}
\end{center}

\index{deblending}
\begin{itemize}
\index{DEBLEND MINCONT}
\item DEBLEND\_MINCONT Minimum contrast parameter for de-blending.
\index{DEBLEND NTHRESH}
\item DEBLEND\_NTHRESH Number of deblending sub-thresholds.
\end{itemize}

\begin{center}
\noindent\fbox{\parbox{0.8\textwidth}{
NOTE! if deblending is too course, a clump of stars can become a 'galaxy'.
}}
\end{center}

\begin{center}
\noindent\fbox{\parbox{0.8\textwidth}{
NOTE! And if deblending is too picky, a single galaxy might be chopped up into  several objects.
}}
\end{center}

There is no single remedy but a DEBLEND\_NTHRESH of 32 and a DEBLEND\_MINCONT of order 0.01 are a good place to start.

\begin{center}
\noindent\fbox{\parbox{0.8\textwidth}{
HMMMM: If the ANALYSIS\_THRESH is different from the DETECT\_THRESH, the 
objects are NOT detected again. So carefull with making these different.
}}
\end{center}

\subsection{Reliability}

This section is one big note of caution. The NNW classification by SE is 
NOT perfect and will break down at the lower magnitude end. Unfortunately it 
does not give another value for unreliable classification (42 for instance)
but assigns a random value between 0 and 1. (see figure \ref{fig:class})

\begin{figure}[htb]
    \centering
    \caption{The dependence of CLASS\_STAR on the luminosity of objects. The reliability clearly disappears at the lower end.} 
    \label{fig:class}
\end{figure}

\clearpage

\section{\label{sec:assocout}ASSOC output}

\index{ASSOC}
\index{ASSOC NAME}
\index{ASSOC TYPE}
\index{ASSOC DATA}
\index{ASSOC }

These are the two output parameters associated with cross correlating catalogs. If you have specified a ASSOC file and the manner of cross-identification (see the input parameters in section \ref{sec:assoc}). 
Be sure to specify one of these as otherwise there will be NO crosscorellation and the SE catalog will contain just as many entries as without the ASSOC parameters specified.\footnote{I will NOT tell anyone how long it took me to figure THAT one out...} The column in the ASSOC\_NAME catalog, specified in ASSOC\_DATA are given in the VECTOR\_ASSOC

\begin{tabular*}{5in}[c]{l p{2.6in} l }

Name & description & unit \\
\hline
\index{VECTOR ASSOC}
 VECTOR\_ASSOC	& ASSOCiated parameter vector  				& \\
\index{NUMBER ASSOC}
 NUMBER\_ASSOC	& Number of ASSOCiated IDs  				& \\
 \hline
\end{tabular*}

\section{\label{sec:flags}Flags}

{\it
How to steal a country through the cunning use of flags:\\
Brittain: I claim India for Brittain!\\
India: You can't claim us! There's 300 million of us already living here! \\
Brittain: Do you have a flag? No flag, no country.\\
--adapted from Eddie Izzard.
}\\

This section is the one in the official manual with some annotations of mine.
\footnote{I steal the official manual's prose from it with reckless abandon, okay?}

A set of both internal and external flags is accessible for each object. 
Internal flags are produced by the various detection and measurement 
processes within SExtractor; they tell for instance if an object is 
saturated or has been truncated at the edge of the image. 

External flags come from flag-maps: these are images with the same 
size as the one where objects are detected, where integer numbers 
can be used to flag some pixels (for instance, bad or noisy pixels).
These types of images are becoming quite common data-products 
of reduction pipelines of current space missions like the Hubble or 
Spitzer telescopes or the grand surveys such as SLOAN.

Different combinations of flags can be applied within the isophotal 
area that defines each object, to produce a unique value that will 
be written to the catalog.

\subsection{\label{sec:flagsint}Internal Flags}

The internal flags are always computed and the default.param file lists them. Clearly  They are accessible through the FLAGS catalog parameter, which is a short integer. FLAGS contains, coded in decimal, all the extraction flags as a sum of powers of 2:

\begin{itemize}
\item[1] The object has neighbours, bright and close enough to signi cantly bias the MAG AUTO
photometry\footnote{This flag can only be activated when MAG\_AUTO magnitudes are requested. }, or bad pixels (more than 10\% of the integrated area affected).
\item[2] The object was originally blended with another one.
\item[4] At least one pixel of the object is saturated (or very close to).
\item[8] The object is truncated (too close to an image boundary).
\item[16] Object's aperture data are incomplete or corrupted.
\item[32] Object's isophotal data are incomplete or corrupted\footnote{An old flag inherited from SExtractor V1.0. It has been kept for compatibility reasons (and why tear it out?). With SExtractor V2.0+, having this flag activated doesn't have any consequence for the extracted parameters.}.
\item[64] A memory overflow occurred during deblending.
\item[128] A memory over flow occurred during extraction.
\end{itemize}

For example, an object close to an image border may have FLAGS = 16, 
and perhaps FLAGS = 8+16+32 = 56. The flags are combined such that 
a unique number will result as the flag entry in the catalogs.

\subsection{\label{sec:flagsext}External Flags}

SExtractor will look for an external flags file when IMAFLAGS\_ISO or NIMAFLAGS\_ISO
are present in the catalog parameter file. The file is specified in FLAG \_IMAGE
The external and internal flags are then combined and listed as the FLAGS output 
parameter in the catalog. So you need to make sure that a combination of flags is 
still unique. There is a fun little exercise for you.

It then looks for a FITS image speci ed by the FLAG IMAGE keyword in the configuration 
file. The FITS image must contain the flag-map, in the form of a 2-dimensional array of 8, 
16 or 32 bits integers. It must have the same size as the image used for detection. Such 
flag-maps can be created using for example the WeightWatcher software (Bertin 1997).\footnote{SEXtractor? Weightwatchers? What's with the naming convention here? I'm almost 
certain that this contributed to my celery-phobia.}

The flag-map values for pixels that coincide with the isophotal area of a given detected object
are then combined, and stored in the catalog as the long integer IMAFLAGS ISO. 5 kinds of
combination can be selected using the FLAG TYPE configuration keyword:
\begin{itemize}
\item[] OR: the result is an arithmetic (bit-to-bit) OR of flag-map pixels.
\item[] AND: the result is an arithmetic (bit-to-bit) AND of non-zero flag-map pixels.
\item[] MIN: the result is the minimum of the (signed) flag-map pixels.
\item[] MAX: the result is the maximum of the (signed) flag-map pixels.
\item[] MOST: the result is the most frequent non-zero flag-map pixel-value.
\end{itemize}

The NIMAFLAGS ISO catalog parameter contains a number of relevant flag-map pixels: 
the number of non-zero flag-map pixels in the case of an OR or AND FLAG\_TYPE, or 
the number of pixels with value IMAFLAGS ISO if the FLAG TYPE is MIN, MAX or MOST.

\section{\label{sec:fitparams}Fitted Parameters}

There is a experimental section of SE which will hopefully become workable soon enough.
This deals with the fitting of the Point Spread Function to stars and light profiles to 
extended objects.In the meantime the follow-up program GALFIT and GIM2D have 
cornered this market. As a result the priority on finishing these parameters may have slipped some...

\subsection{\label{sec:psf}PSF fitting}

\index{PSF}
\index{point spread function}
The point spread function of an instrument describes how the light from a 
point source is distributed over the detection element of a CCD. If you 
have a little fits file of a model of the PSF and you've figured out 
how to feed that to SE (if you do let me know...) then a part of SE known as 
PSFeX can fit this to every object. The following parameters should then 
be available to you. As you can see some of these are either photometric, 
geometric or positional parameters.
I've just list them here as this is still an experimental bit of SE but 
should be very useful later.\\ 
The PSF is also used for fitting the components of a galaxy in the next 
section.

\begin{tabular*}{5in}[c]{l p{2.6in} l }

Name & description & unit \\
\hline
\index{XPSF IMAGE}
  XPSF\_IMAGE	&	X coordinate from PSF-fitting			& pixel  \\
\index{YPSF IMAGE}
  YPSF\_IMAGE	&	Y coordinate from PSF-fitting			& pixel  \\
\index{XPSF WORLD}
  XPSF\_WORLD	&	PSF position along world x axis			& deg  \\
\index{YPSF WORLD}
  YPSF\_WORLD	&	PSF position along world y axis			& deg  \\

\index{ALPHAPSF SKY}
 ALPHAPSF\_SKY	& Right ascension of the fitted PSF (native)	& deg  \\
\index{DELTAPSF SKY}
 DELTAPSF\_SKY	& Declination of the fitted PSF (native)	& deg  \\

\index{ALPHAPSF J2000}
 ALPHAPSF\_J2000	& Right ascension of the fitted PSF (J2000)	& deg  \\
\index{DELTAPSF J2000}
 DELTAPSF\_J2000	& Declination of the fitted PSF (J2000)		& deg  \\

\index{ALPHAPSF B1950}
 ALPHAPSF\_B1950	& Right ascension of the fitted PSF (B1950)	& deg  \\
\index{DELTAPSF B1950}
 DELTAPSF\_B1950	& Declination of the fitted PSF (B1950)		& deg  \\

\index{FLUX PSF}
 FLUX\_PSF	& Flux from PSF-fitting				& count  \\
\index{FLUXERR PSF}
 FLUXERR\_PSF	& RMS flux error for PSF-fitting		& count  \\
\index{MAG PSF}
 MAG\_PSF	& Magnitude from PSF-fitting			& mag  \\
\index{MAGERR PSF}
 MAGERR\_PSF	& RMS magnitude error from PSF-fitting		& mag  \\

\index{NITER PSF}
 NITER\_PSF	& Number of iterations for PSF-fitting  	& -  \\
\index{CHI2 PSF}
 CHI2\_PSF	& Reduced chi2 from PSF-fitting  		& -  \\
\hline
\end{tabular*}

\noindent All these have ERR+parameter uncertainties associated with them, except NITER\_PSF and CHI2\_PSF of course. That is...if these ever work...

The reason you'd want a decent fit of the PSF to an object is to do really good photometry on it 
as you're now accounting for al that leaked light. Note that such a correction would only work if
the object was originally a pointsource. This is fairly useless for extended objects but for that we have the galaxy fitting parameters.

\subsection{Galaxy profile fitting}

Another experimental bit of SE. As you might know, average radial lightprofile of a galaxy 
can be described by fit. In the case of a spiral galaxy it is an exponential for the disk 
and a deVaucouleur ($r^{-4}$) profile for the bulge. The ($r^-{4}$) profile fits an elliptical 
galaxy quite well. So if the galaxy is reasonably resolved, this new bit of SE can fit these 
profiles. It can give you the bulge disk ratio's and everything. Provided it'd actually work.\footnote{As per v2.4.4 they do not...}
Did I mention GALFIT and GIM2D in this context?

\begin{tabular*}{5in}[c]{l p{2.6in} l }
Name & description & unit \\
\hline
\index{FLUX GALFIT}
 FLUX\_GALFIT	 & Flux derived from the galaxy fit			& count \\  
\index{FLUXERR GALFIT} 
 FLUXERR GALFIT	 & RMS error for GALFIT flux			 	& count \\  
\index{MAG GALFIT}
 MAG\_GALFIT	 & Magnitude derived from galaxy fit			& mag  \\ 
\index{MAGERR GALFIT}
 MAGERR\_GALFIT	 & Magnitude error derived from galaxy fit		& mag \\  
\index{ERROR GALFIT} 
 ERROR\_GALFIT	 & Reduced Chi-square error of the galaxy 		& fit  \\
\index{GALDANG IMAGE}
 GALDANG\_IMAGE	 & Galaxy disk position angle  from the galaxy fit 	& deg  \\ 
\index{GALDSCALE IMAGE} 
 GALDSCALE\_IMAGE & Galaxy disk-scale from the galaxy fit   		& pixel \\  
\index{GALDASPEC IMAGE} 
 GALDASPEC\_IMAGE & Galaxy disk aspect ratio from the galaxy fit 	& \\
\index{GALDE1 IMAGE} 
 GALDE1\_IMAGE	 & Galaxy disk ellipticity nr1 from the galaxy fit 	&  \\
\index{GALDE2 IMAGE} 
 GALDE2\_IMAGE	 & Galaxy disk ellipticity nr2 from the galaxy fit 	&  \\
\index{GALBRATIO IMAGE} 
 GALBRATIO\_IMAGE & Galaxy bulge ratio from the galaxy fit  		& \\
\index{GALBANG IMAGE}
 GALBANG\_IMAGE	 & Galaxy bulge position angle  from the galaxy fit 	& deg   \\
\index{GALBSCALE IMAGE} 
 GALBSCALE\_IMAGE & Galaxy bulge-scale from the galaxy fit		& pixel \\  
\index{GALBASPEC IMAGE} 
 GALBASPEC\_IMAGE & Galaxy bulge aspect ratio from the galaxy fit 	&  \\

\hline
\end{tabular*}

\section{Principle Component}

Again these are related to the psf fitting SE is -hopefully- capable of in the near \footnote{Meaning: 'possibly in your lifetime'.} future. \\

\begin{tabular}[c]{l l l }
Name & description & unit \\
\hline
  X2PC\_IMAGE	&	PC variance along x				& pixel$^2$   \\
  Y2PC\_IMAGE	&	PC variance along y				& pixel$^2$   \\
  XYPC\_IMAGE	&	PC covariance between x and y			& pixel$^2$   \\
 
  APC\_IMAGE	&	PC profile RMS along major axis			& pixel   \\
  BPC\_IMAGE	&	PC profile RMS along minor axis			& pixel   \\
  THETAPC\_IMAGE	&	PC position angle (CCW/x)			& deg   \\
  PC		&	Principal components	  			& \\
\hline
\end{tabular}

\chapter{Strategies for SE use}

\begin{center}
{\it Baldrick: "Sir, I have a cunning plan"\\
Blackadder: "As cunning as a fox who is professor of cunning at Oxford University?"\\
}
\end{center}

This section is reserved for some lecturing on what I found were 
good tricks to use with SE. There are some tricks you can do to 
extend the depth of your exposures, to make it easier to use 
SE on a batch of objects or tune your detections more to your needs.

%
%
%
%

\section{Image types to use?}

There are several options open to you for the image type. It is quite usual 
for astronomical images to be the result of stacked exposures. There are 
several ways to stack. The interaction between the type of addition and the 
photometric inputparameters is discussed here.

You may want to ask yourself whether you want to use the original integrated 
image with the total counts of every object or to use these divided by the 
exposure time, essentially counts-per-second images.
Using counts-per-second images is not as dumb as it sounds. 
By using counts-per-second images, the dynamic range is the same for 
images with wildly varying exposure times. When examining them, it makes 
them much easier to compare. The signal-to-noise ratio (S/N) is of 
course still the same.

In the case of counts-per-second images, you can determine the 
MAG\_ZEROPOINT for an 1 sec exposure and put this in the parameter file.
Only the GAIN varies with every different exposure. The value for the GAIN 
in this case is 

\begin{tabular*}{5in}[l]{p{1.7in} p{1.4in} p{1.7in}}
Effective Gain 		   & Magnitude zeropoint & Type of image \\
\hline
gain $\times$ total exposure time & zeropoint(1 \ sec) & input image is c/s \\
gain 			   & zeropoint(1 \ sec) + 2.5 log$_{10}$(exp. time) & sum of N frames \\
N$\times$gain 			   & zeropoint(1 sec) + 2.5 log$_{10}$(av. exp. time) & average of N frames \\
2$\times$N$\times$gain/3 		   & zeropoint(1 sec) + 2.5 log$_{10}$(av. exp. time) & median of N frames \\
\hline
\end{tabular*}\\

the ccd gain times the exposure time.
\footnote{As you've probably guessed I used counts-per-second images. 
It's personal preference but it's much easier to run SE in batches if 
the only thing you have to change is the GAIN. The zeropoint changes as 
soon you change filter of use a different ccd in an array like changing 
from WF2 to WF3 of the Wide Field camera on Hubble.}

\subsection{Thresholds}

This last sheme works best when your thresholds are RMS noise related. 
Otherwise you'd have to modify the thresholds anyway to account for 
the signal-to-noise. The zeropoints of the thresholds remain the same 
for counts-per-second images and integrated images. This can be {\it very} 
confusing, using the zeropoint-for-one-sec-exposure in the MAG\_ZEROPOINT 
parameter and the zeropoint-for-one-sec-exposure + 2.5log(exptime) as the 
zeropoint for the detection/analyse threshold.

\section{How to get faint objects?}

Faint objects are the ahrdest and often the most interesting objects in you 
image. So how to get them? There are tricks for the detection images, the thresholds and the filters you use. 

\subsection{Different Detection Images}

There are several strategies employed to detect and classify sources, using 
Source Extractor. Source Extractor has the in-built ability to detect sources
 in one image and subsequently do photometry on the found apertures in 
another image. This has the advantage that the photometry on an object 
has the same apertures in all bands and the catalogs need not be matched
(the numbering for all the catalogs is the same).
Many schemes use this ability.

\subsubsection{``Meta''-images}

One idea is to add data out of two or several bands and do the  
detection there while doing the photometry in the actual data. The image 
where the detection is done is referred to as a "meta-image"
(i.e. \cite{1996AJ....112.1335W} used an I+V image to detect sources in 
the HDF). This is quite widespread use; you can use a median, mean or a stacked images. Just beware of differences in PSF when making these images.
\footnote{The PSF of the resulting image should be about that of the worst PSF component. Some convolution may be needed.}

 The advantages are that 
apertures can be determined more accurate for objects, espescially 
faint ones. However, strucural parameters (all those computed from the 
object's moments) are determined for the meta-image. These may not always 
be very representative of your objects.

\subsubsection{Maximum likelyhood images}

Another approach using multi-band images is to construct a $\chi^2$ image 
and do the detection on these. \cite{1999AJ....117...68S} introduced this 
technique on the HDF. However this method appears to be most successful in 
the truly multiband (more than three at least) exposures.

Both of these are aimed at getting as faint as possible sources.

\subsubsection{Optimal image}

Most commonly, the band where your target objects are the brightest
(i.e. the reddest band available for field galaxies) or the band 
with the longest exposure is used as the detection image and the apertures 
of that band are used for the photometry in all the other bands.
The detection of background galaxies is done in the band with 
the longest wavelength available is used on the premises that background 
galaxies appear red.

Alternatively SE can be run on each band separately and the catalogs 
are subsequently matches by position. Fixed aperture photometry, 
corrected for the PSF can then be used to avoid differences due to different 
apertures. This had obvious drawbacks for extended or faint sources; you 
don't know how well the apertures match.

\subsection{Thresholds and Filters}

Again the settings within sextractor can be optimized for faint objects. 
The threshold can be set as low as you want it but a DETECT\_THRESH of 1 
$\sigma$ RMS above the background is probably the best you can do. However 
how that background is determined is very imposrtant. Depending on your 
images and objects, changing the BACK\_SIZE background mesh size can improve 
your detection of faint stuff, especially near bright objects (v hard in any 
case). Which brings us to the CLEAN option you might want to turn off so you 
can better detect faint stuff near brighter things. 

And some of the filters used to smoothe the image before detection are 
influential in the faint sources detection. Have look at the filtering 
section earlier. Usually a Gauss approximating the PSF in size is used. 

\begin{center}
\noindent\fbox{\parbox{0.8\textwidth}{
NOTE!; whatever your settings, MAG\_AUTO is the best estimate of the magnitude but it still underestimates by as much as several tens of magnitude for faint objects. \footnote{I have not experienced this myself but it was reported from several sides, especially for faint galaxies. Some of the flux simply still is from outside the Kron radius. The correction did not look very linear.}\\
}}
\end{center}

Best way to figure this out is to inject you images with simulated objects of 
known magnitude and check their magnitude in the SE catalog.
For simulated object, I use {\it mkobject} in IRAF. Give it a list of 
positions, magnitudes and profiles and you've got your calibration objects.

\section{How to get good colours of objects?}

Colours of objects found in surveys are a very important tool for astronomers.
So having a good strategy how to get the colours from SE catalogs is probably 
a good idea. There are several ways to determine these. Do you use a detection
 image and photometry image(s) or use values from several independent catalogs?
Which magnitudes to use;ISO, ISOCOR, MAG\_AUTO, MAG\_APER or something else?

\subsection{separate detection and photometry images?}

SE has this ability to use the apertures from one detection image in the other, photometry image. This should give accurate colours as the apertures the same right? Some things to consider:

\begin{itemize}
\item The images better be aligned to the pixel. Small shifts can result in dramatically different colours.
\item The PSF's and seeing better be similar (i.e. the pixels in the detection image are indeed the same part of the object in the photometry image)
\end{itemize}

\subsection{which output to use?}

There is only two to consider MAG\_ISO and MAG\_APER, the others are too dependent on the SE settings in the detection.

\subsubsection{MAG\_ISO}
The best aperture for colours is the MAG\_ISO. This gives the flux in an as big an aperture that will fit in the object, will be roughly the same shape as the object and when using separate detection and photometry images, these will give you the most accurate colours. This of course does not hold of you crosscorellate catalogs.

\subsubsection{MAG\_APER}
MAG\_APER seems ideal for the colour of objects; either used in separate detection and photometry images or from crosscorrellated catalogs, the apertures are user specified and comparable over the bands used (you can correct the radii for instance with seeing...) BUT! there are a few things to consider with fixed apertures:

\begin{itemize}
\item crowding; does your aperture overlap with another object?
\item objects smaller than the aperture will have less reliable colours
\item Still, are they aligned? Is this the same part of the object?
\item In the case of small apertures, is the colour of the center of the 
object the same as the for the whole? 
\item PSF aperture corrected?
\end{itemize}

But you can take these in account. The only (dis)advantage
\footnote{good or bad? your call.} MAG\_APER has is 
that really the fact that it is fixed for every object. However it is not so  
dependent on the detection by sextractor. \\

In conclusion: 
with detection and photometry images, use MAG\_ISO, with separate detections, use MAG\_APER. Ponder which ones would be best for you type of objects.

\section{Finding your objects of interest}

This is a very broad query. Literate searches and some intense staring at 
typical objects should give you an idea what the characteristic values of 
SE output of your objects are. Sometimes a specialised parameter added to 
SE might be able to help or a ratio of SE parameters (aperture fluxes for 
instance). Best thing to do is get a trainingset and start plotting. 
FWHM, ELLIPTICITY, MAG\_APER(S) FLUX\_RADIUS and concentration (MU\_MAX over MAG\_AUTO is a reasonable indication but there are more) are good places to start.

And of course colours always help.

\section{Strategies to find galaxies in crowded fields}

To increase SE's ability to find field galaxies in crowded fields, several 
strategies may be considered.

\cite{Gonzalez} used B-I images of NGC 3664 to detect sources instead of 
the I images.
This has the advantage that stars can be negated somewhat, leaving only objects of 
a severely differing color. The disadvantage is the increase in the noise, 
increasing the chances of spurious detections. And there actually has to be 
a B band exposure.

An alternative approach is to remove the stars, using DAOPHOT. It is honed
 to find and model stars in crowded fields and there is the option to 
produce an image with the model subtracted. Again, this could increase 
the number of spurious detections because of noise.

Alternatively SE can simply be run on the I images and photometry can be 
done on the V image. This additional information can be used to filter the 
catalogs for actual galaxies. As an additional filter, an visual 
inspection of the data can be employed.

A technique to enhance faint extended sources is called unsharp masking. 
Originally developed for photographic plate, ccd images would be smoothed 
and the smoothed version would be subtracted from the original. This 
technique works fine for filamentary structures like shells and tidal 
tails but is unsuited for finding galaxies in crowded fields. (SE does 
however smooth the image slightly and then does the detection)

Metaimages, the addition of two band, suitably weighed, as detection 
images is another popular option. Again this opens the possibility of 
many spurious detections.

\chapter{Available Packages}

Maybe it's not a good idea to advertise the {\it competing} software but 
SE might not exactly be what you need. Better if you figure this early.
At the moment there are several programs and approaches in use that 
attempt to detect and classify objects; 
Sextractor \cite{SE}, FOCAS \cite{FOCAS,FOCAS2}, DAOPHOT \cite{DAOPHOT} 
and self organising maps (SOM),  

All of these have been tried and tested in relatively uncrowded fields 
and are expected to break down to some extent in crowded fields. The 
notable exception is DAOPHOT which is honed on crowded stellar fields. 
However it is focused on accurate stellar photometry, not on detecting 
extended objects.

FOCAS is widely rumored not to work, especially in crowded environments
 but still popular. I personally haven't used it but SE is somewhat more recent
so I chose SE over FOCAS. 

Self Organizing Maps (SOM) are still in an experimental phase and have 
proved themselves superior to SE in classifying galaxies by using an 
additional neural network. However they are not distributed in a package. 
If you {\it really need} detailed classification of objects (your thesis 
hangs on it being Sab galaxies) then you'd better contact the authors of 
papers on SOMs or write something yourself. But if you can think up a reasonable 
parameter that will help you classify, you can always define your own in SE (see 

 This leaves however SE as the current program 
of choice, almost by default.

\chapter{\label{ch:follow-up}Follow-up Programs}

SE is now used as either a quick look or -mostly- as the first step in a more involved 
analysis that often involves fits to the images. In the case of galaxies, there are several 
packages to model the light distribution and determine interesting fitted parameters.
There seem to galaxy fitting parameters (at least) out there, GALFIT and GIM2D. 
I honestly cannot tell you which one is better.

\section{\label{sec:GIM2D} GIM2D}

GIM2D is a program that analyses galaxy images and spits out a galaxy-subtracted 
image (good to check the model with) and parameters. The use of SE is a crucial 
step in the process. 

website:
\begin{verbatim}
http://www.hia-iha.nrc-cnrc.gc.ca/STAFF/lsd/gim2d/
\end{verbatim}

\section{\label{sec:GALFIT} GALFIT}

GALFIT also fits galaxy images. This was not initially intended to be used in automated 
batch analysis and SE is not a crucial step in the process. 

website:
\begin{verbatim}
http://zwicky.as.arizona.edu/~cyp/work/galfit/galfit.html
\end{verbatim}


\chapter{SE use in the Literature}

In the literature SE is used for a myriad of problems. In this section 
the usage for detecting galaxies in the literature is discussed. It is 
not up to date completely, lots of people continue to use SE. Keep a 
look out in the literature.

\cite{Casertano2000} used WFPC2 (U B V I) and a V+I detection image of the 
HDF-S with the following settings: THRESH 0.65 $\sigma$,MIN\_AREA 16 
(0.05'' pixelsize) and MIN\_CONTAST 0.03 . They used the weight maps made by 
drizzle for RMS background estimate.\\

\cite{Brown2000} used SE on photographic plates (UBRI) and made separate 
catalogs for each band. Classified as star if class $>$ 0.7 in three bands 
or class $>$ 0.75 in two bands or class $>$ 0.85 in one band.

\cite{Rodighiero2000} used SE on the HDF-N plus JHK band. detection was in 
the K-band excluding Vaucouleur-profiles. FLUX\_RADIUS was used in photometry 
correction.

\cite{Smail2001} used SE on WFPC2 BVI with THRESH 1.5 $\sigma$ 21.2 mag 
arcsec$^{-2}$ and MIN\_AREA 10. Rejection criterion: k < 19 within 1-2 
half light radii.

\cite{Kalirai2001} CFHT BVR data. Has a plot of magnitude -stellarity with 
a cut for the best star-galaxy separation. Mention of PSFex mentioned for 
correction of photometry.

\cite{Williams2001} run SE on KeckII data (BVI) to V = 26.5. All objects 
with V<24 and CLASS < 0.2 is a galaxy.

\cite{Rodighiero2001} run SE same as \cite{Franceschini1998}. SE is used 
to perform simultaneous slicing. From this, the PSF and the profiles for 
all objects are constructed. Galaxies are then selected as those objects 
resembling the Veaucouleurs profile convolved with the PSF...
This is done for morphological classification.

\cite{Trujillo2001} used BVRI data from the NOT. I band detection with 
LOWTHRESHOLD 1.5$\sigma$ MINAREA 4. 2.5 kron radius photometry. 

\cite{Vaisanen2000} used SE on J,K data from the 1.2 m telescope. Used 
MAG\_BEST and eyeball identification.

\cite{Zabludoff2000} used SE in a standard config on rosat's PSPC 
considering everything with CLASS $>$ 0.5 as a galaxy. Checked with 
isophotal area. MAG\_BEST photometry.

\cite{Kambas2000} used ground based data with MINAREA 5 (2''.3 pixelsize)
and THRESH 1 $\sigma$. The SE catalog was then filtered for flagged 
objects. VLSB sample selection: $\mu_0$ $>$ 23 R $\alpha$ $>$ 3'' isophotal 
area $>$ 148 arcsec$^2$. 

\cite{Hogg2000} used SE separately on UGR and K images. the FWHM of the 
smoothing Gauss was picked equal to the seeing. THRESH 1.2 $\sigma$ 
and MIN\_CONTAST 0.01 There was no Star/Galaxy separation.

\cite{Volonteri2000a} used SE on the data of HDF (UBVI) south, in separate 
detections. The DEBLEND\_MINCON 0.01 DEBLEND\_NTHRESH 32 smoothing Gauss 
0.16 arcsec and MINAREA is equal to the seeing disk (!?!) THRESH 1.34. 
Photometry was ISOPHOT\_CORR or aperture, depending on diameter ISO (1.2 
arcsec limit). Stars have class $>$ 0.9 and I $<$ 22

\cite{Volonteri2000b} used SE separately on the UVBI data  of the HDF-S and later on a 'meta-image' of all filters combined. 
with THRESH 1.34 $\sigma$ and 
MIN\_AREA 13 (seeing disk of $\approx$ arcsec$^2$. Photometry was either 
ISOPHOT or aperture depending on isophotal diameter. The limit was 1.2 
arcsec. To estimate the {\it number} of spurious sources, the image was 
inverted and SE was run again on this image detecting only spurious sources. 

\cite{Sowards2000} give no details on their use of SE except that UBVRI 
detections were done separate and photometry was done on 20'' apertures.

\cite{Lubin2000} used a $\chi^2$ image for the detection. This was made 
from three bands (BRi)using the Hale 5m telescope. 

\cite{Szalay1999} proposed the $\chi^2$-images as the detection image. 
Seems only applicable when there is true multicolor information.

\cite{Broadhurst2000} give absolutely no details except that stars are 
badly fit by a redshifted galaxy spectra (Well DUH!)

\cite{Castander1999}used SE to find candidates for visual inspection.

\cite{Gebhardt1999} used SE to search for Globular clusters in HST (VI) data.
Globular clusters are those objects with an mag\_err $<$ 0.1, V-I between 
0 and 2, ellipticity $<$ 0.5, FWHM between 1 and 4 pixels and 'nonstar' 
classification. They estimate a constant contamination from background 
galaxies.

\cite{Poli1999} use a 'meta-image' from BVRI images. UBVRIJK photometry 
was determined from I=23.25 isophotes and 2''.2 and 5'' apertures. Half 
light radius, z and Magnitude relations. Not very useful for galaxy id 
however.

\cite{Treu1999} used SE on NICMOS, WFPC and ground data. They distinguished 
between resolved and unresolved object by comparing with stellar FWHM's. 
The limiting magnitude was the bin where all objects had a MAGERR\_BEST 
$<$ 0.15.

\cite{Menanteau1999} used SE on HST archival data for spheroidal galaxies.
They used both visual identification and the concentration and asymmetry 
parameter from \cite{Abraham1996, Abraham1994}. They plot how to separate 
spirals from ellipticals using these parameters.
 
\cite{Hashimoto1999,Hashimoto1998} used this concentration parameter to 
discern between early and late type galaxies.

\cite{Simard1999} used Source Extractor on the Groth survey strip, THRESH 
1.5 $\sigma$, MIN\_AREA 10 and then fitting the light profiles of found 
galaxies.

\cite{Nonino1999} describe the SE program in detail, especially the detection 
process. The star galaxy separation is less stringent and said to favor 
complete star catalogs at class $>$ 0.5 and more complete galaxy 
catalogs at class $<$ 0.75. They present a flag list of SE. Parameters 
used:BACK\_SIZE 64 (17''), 
THRESH 0.6 $\sigma$. They use RMS images as weight maps (made by Weight 
watchers routine). MAG\_AUTO is preferred and it is explained why.
Good paper on the behavior of SE.

\cite{vanDokkum2000} touches only lightly on the use of SE but describes a 
CLEANing method to improve the resolution of HST/WFPC data. 

The study of high-z clusters by \cite{Oke1998} used FOCAS, SE and MDS 
'find' algorithm. SE is used on the K' band images from the IRIM camera at 
the 4m Kitt peak. THRESH 1.5 $\sigma$, ($\mu_{K'}$ = 22.2 arcsec$^2$), 7x7 
top-hat filter, MIN\_AREA 35(0''.15) The authors conclude that FOCAS and 
SE do not show significant differences in photometry, astrometry of 
classification).

\cite{Hilker1999} present ground-based data and use SE in the V-band.. 
The background mesh is either 56x56 or 128x128 FWHM of the convolving 
gauss is either 1''.5 or 2'' MIN\_AREA 5. The CLASS identification is 
taken to be accurate up to V $<$ 21. The limit was taken to be 0.35 up 
to V = 22. Color is used to identify cluster membership.

\cite{Smail1998} naively used the MAG\_BEST but used 1'' diameter apertures
 on WFPC data and 3'' on ground based data for galaxy colors. In the cases
 where Iband data was not available R band data was converted assuming R-I 
$\approx 0.5 \pm 0.2$

\cite{Teplitz1998}  used SE on NICMOS data with MIN\_AREA 6 and used the 
FWHM of object, provided the were bright enough for a reliable estimate, 
to resolve stars from galaxies. All fainter objects were assumed to be 
galaxies.

\cite{Marleau1998} use SE for the initial detection of galaxies in the HDF
and then determine a string of parameters with GIM2D to quantify the 
morphology. THRESH 1.5$\sigma$, MIN\_AREA 30, DEBLEND\_MINCON 0.001.

\cite{Yan1998} used NICMOS data FWHM 0.3'' and a 2$\sigma$ detection and 
a 1$\sigma$ analysis threshold. Plot of half-light radius vs H magnitude 
with stars and galaxies.

\cite{Brown2001} analyze UBRI data from photographic plates from the SGP 
and UK Schmidt field. No details.

\cite{Fasano1998a} used SE on the HDF to find early type galaxies. THRESH 
1.3 $\sigma$. The selectionlimits for a galaxy were: $V_{606}(STMAG) < 
26.5 \ \  N_{pix} \geq 200 \ \  CLASS \leq 0.6$. \cite{Fasano1998b} used 
this data to establish a SB- effective radius relation for the early type galaxies. Note that they chose to use the V band for identification due to better S/N.

\cite{Arnouts1999} used their own program to make a survey of southern hemisphere faint galaxies. More information than surmisable here.

\cite{Hogg1997} use SE on U and R band images. No specifics are given except that the detection was done on images smoothed with the PSF.

\cite{Bertin1997} used SE on digitized Schmidt plates. It contains a good description of the usage of SE on photographic plates and the particular problems encountered.

\cite{Gardner1996} use SE on BVIand K band ground based images. They use SE on all of them and note that I-K and B-I color are good separators for star galaxy for all except the bluest objects. All identified galaxies up to certain limits were confirmed by eye.

\cite{Lanzetta1996} use SE on HST/HDF images. Detection is in the I band. 
The FWHM smoothing was 0.12 arcsec (approx width PSF) to aid detection of faint sources. Finally there is a reference as to {\it why} you should bother to smooth again with approximately the PSF: \cite{Irwin1985}.MIN\_AREA 10 THRESH 1.4 $\sigma$ and surprisingly a CLEAN\_PARAM pf 2.0 to get rid of closely-packed objects (taken to be one object) BACK\_SIZE was an unusual 41x41 pixels. As detection was done in the I band image, photometry was done in dual mode for the rest of the bands.

\cite{Smail1997} produced catalogs on 10 clusters images by HST. Detections as in the reddest band available (either F702W or F814W). The modified SE code to produce the concentration index introduced by \cite{Abraham1994, Abraham1996} and the contrast index which measures the fraction of light in the brightest 30\% of its pixels. Object were however checked visually. THRESH 1.3 $\sigma$ is the only parameter mentioned.

\chapter{Known Bugs and Features}

\begin{figure}[htb]
    \centering
    \caption{Comic taken from http://www.phdcomics.com/. No royalties paid. Hope that's ok. Please don't hurt me.}
\label{fig:bug}
\end{figure}

SE was, still is, always will be in develoment. After all it's pretty much one guy coding it all up (Emmanuel Bartin if you've missed that). So there are still outstanding issues. Here are the ones I found. They'll be mentioned throughout but here's the list:

\begin{itemize}
\item[1.] The PSF fitting routine does not work (yet?).
\item[2.] Neither does the galfitting routine. But hey there are follow-up scripts for those.
\item[3.] The ELLIPTICITY and all the moments is determined from the smoothed image. This is a faeture not a bug but it is not what people intuitively expect SE to do.
\item[4.] The first pixel's coordinates is 1,1. If you're thinking in terms of matrices, this can be annoying.
\item[5.] The documentation will always lag the program (but to claim this is unique to SE...)
\item[6.] Output parameters asked for are important. This is an SE feature. If no ASSOC output parameter is asked for, the assoc module is not run. If no Kron-dependent parameter is in the param file, this aperture is not drawn on the APERTURE image. It's a feature you need to be aware of.
\end{itemize}

\chapter{SE parameter additions}

\index{adding parameters SE}
\index{parameter addition}
\index{new parameters}	
It's also discussed in the manual but there is a possibility that the user can define his/her 
own output parameters in SE. In order to do that, you need to do the following things: 
First ask yourself if you can't do it with ANY of the parameters above or combinations thereof.
 After all there is a lot of them and it a lesser headache.\\
If not, then define what this parameter should be and then modify the analyze.c, types.h and 
param.h as follows:
Define your parameter after the definition of the FWHM function in analyse.c 
(after line 371 in version 2.2.2). Define the parameter name in types.h and param.h like so:\\
In types.h
\begin{verbatim}
  float         conc;       /* IRS concentration index */
\end{verbatim}
and in param.h:
\begin{verbatim}
  {"CONCENTRATION", "Abraham concentration parameter",
        &outobj.conc, H_FLOAT, T_FLOAT, "%8.3f",""},
\end{verbatim}

This way, if you include CONCENTRATION in the file given to PARAMETERS\_NAME, then the concentration will be calculated and put in the value 'conc'. \\

This parameter is the concentration parameter as defined in 
\cite{Abraham1994,Abraham1996} and previously implemented by Ian Smail in \cite{1997ApJ...490..577D
,1998ApJ...497..188C}. I've modified it so that it works in version 2.2.2.

The contrast and asymmetry parameters are given as examples later on.

\begin{verbatim}  
  /* Abrahams concentration index calculated - IRS/BWH */
  /* fraction of light in central 30% of the objects area, measured 
  in an ellipse aligned with the object and having the same axis ratio */
  if (FLAG(obj.conc)) 
    {
    double xm,ym,dx,dy,rv,cv,amp; # local variables
    double AA,BB,CC,DD,EE,FF,sintheta,costheta ; 
    
    xm = obj->mx; # get the median x position from the object struct
    ym = obj->my; # get the median y position from the object struct
    amp = tv/(2*PI*obj->a*obj->b*obj->abcor);
    rv = cv = 0.0;
    for (pixt=pixel+obj->firstpix;pixt>=pixel;pixt=pixel+PLIST(pixt,nextpix))
      # run through all the pixels in the object struct
      {
	dx = PLIST(pixt,x) - xm; # calculate dx and dy
	dy = PLIST(pixt,y) - ym;
	pix = PLIST(pixt,value)<prefs.satur_level?
	  PLIST(pixt,value)
	  : amp * exp(-0.5*(obj->cxx*dx*dx + obj->cyy*dy*dy + obj->cxy*dx*dy)
		      /obj->abcor);
	
	costheta= cos(obj->theta*PI/180.) ;
	sintheta= sin(obj->theta*PI/180.) ;
	EE = (obj->b>0.0)?sqrt(obj->npix*obj->a/(PI*obj->b)):1.0;
	FF = (obj->a>0.0)?sqrt(obj->npix*obj->b/(PI*obj->a)):1.0;
	
	AA= costheta*costheta/EE + sintheta*sintheta/FF ;
	BB= CC= sintheta*costheta*(1.0/EE - 1.0/FF) ;
	DD= + sintheta*sintheta/EE + costheta*costheta/FF ;
	if ((AA*dx+BB*dy)*(AA*dx+BB*dy)+(CC*dx+DD*dy)*(CC*dx+DD*dy) < 0.09)
	  cv += pix;
	if ((AA*dx+BB*dy)*(AA*dx+BB*dy)+(CC*dx+DD*dy)*(CC*dx+DD*dy) < 1.00)
	  rv += pix;
      }
    obj->conc = (rv>0.0)? (cv/rv): 99.0;
    }

\end{verbatim}

\chapter{\label{ch:examples}Examples}


In this chapter I intend to show the workings of Source Extractor using the 
Hubble Deep Field and Hubble Ultra Deep Field. Of course this does not 
cover the many many uses of SE but it should give you an idea.

\section{\label{sec:hdf}The Data: the Hubble Deep fields}

The Hubble deep fields were astronomy's first look at the high-redshift universe. 
As a result these fields have been analysed to death and are well known. 
These also saw some of the first quick-look papers that almost always used 
SE for the statistics. Also, high-end science products are freely available. 

The Hubble Deep Field (North) can be found here:\\
\begin{verbatim}
http://www.stsci.edu/ftp/science/hdf/hdf.html
http://stdatu.stsci.edu/hst/hdf/v2/mosaics/x4096/
\end{verbatim}
I am using the files which can be found here:
\begin{verbatim}

\end{verbatim}

The Hubble Ultra Deep field is a Advanced Camera for Surveys product. 
This is a much bigger field and hence file. I'll use these to illustrate the 
capability of SE to habdle larger files\footnote{I realise that computers 
are getting bigger and better so the definition of ``a large file'' is constantly 
changing but the UDF is freely available and easy to obtain.}

\begin{verbatim}
http://www.stsci.edu/hst/udf
http://archive.stsci.edu/prepds/udf/udf_hlsp.html
\end{verbatim}

\section{\label{ex:default}Running the SE default}

Simpelest place to start. Put the default files in the config directory from 
source extractor in the same directory as the f814\_mosaic\_blk4.fits file. Run SE with:

\begin{verbatim}
>  sex f814_mosaic_blk4.fits
\end{verbatim}

And that is it! This has fathered\footnote{sex, fathered...get it? Oh the puns...ugh. 
} a catalog named `test.cat'.

Since the default.param output was generated this contains the following parameters:

\begin{verbatim}
#   1 NUMBER          		Running object number
#   2 FLUXERR_ISO     	RMS error for isophotal flux   [count]
#   3 FLUX_AUTO       	Flux within a Kron-like 
					elliptical aperture                     [count]
#   4 FLUXERR_AUTO    	RMS error for AUTO flux         [count]
#   5 X_IMAGE         		Object position along x           [pixel]
#   6 Y_IMAGE         		Object position along y           [pixel]
#   7 FLAGS           		Extraction flags
\end{verbatim}

And 1714 entries. The checkimages can be obtained as well by running this:

\begin{verbatim}
sex f814_mosaic_blk4.fits -checkimage_type APERTURES,BACKGROUND 
-checkimage_name aper.fits,back.fits
\end{verbatim}

these are in figure \ref{fig:hdf}. 

\begin{figure}[htb]
    \centering
    \caption{The Hubble Deep Field North (F814W) and the various checkimages when SE is run with the default settings. The apertures file shows the KRON apertures around the objects. }
\label{fig:hdf}
\end{figure}

\section{\label{ex:star/gal}Stars and Galaxies}

Now we go into the default.param file and comment out everything 
except the MAG\_AUTO and the CLASS\_STAR parameters. 
If we plot these two values (still using default values for the 
source extraction) we get figure \ref{fig:hdf:class}. 
 
\begin{figure}[htb]
    \centering
    \caption{The MAG\_AUTO and CLASS\_STAR parameters of the Hubble Deep Field. The MAG\_APER has strange values as the zeropoint setting is wrong. Most of the objects are classified as galaxies but the classification becomes random for the dimmer objects.  }
\label{fig:hfd:class}
\end{figure}

Note how similar it looks to figure \ref{fig:class}. What is you need to 
take away is that the star/galaxy separator is okay to give an indication 
but that it becomes a random number genereator pretty quick.
There are not that many stars in the HDF yet there are quite a number 
of objects classified as such. As with most classification schemes, SE 
finds out if an object is extended or not. Not extended, must be a star...

\section{\label{ex:phot}Photometry}

\section{\label{ex:radii}Typical Radii, source sizes}

\chapter{\label{ch:ack}Acknowledgments}

I have had help with SE over the years. Asked people what they thought of the 
program or their questons helped me. Oh and the people who kept paying me 
my wages even though I spent my time working on this manual.
I would like to thank Ian Smail, Ernst-Jan de Vries, Anton Koekemoer, Ed Smith, 
Emmanuel Bertin, Harry Ferguson, Roza Gonz\'alez and Ron Allen.
Lauren Grodnick was kind enough to help with language issues of the earlier versions.

\appendix

\chapter{'Drizzle' and RMS weight images}

Suppose you're a HST/WFPc user and you want to use SE on the data. The program 
'drizzle' which combines a series of exposures produces also weight maps of the 
'drizzling'. These weight maps can be used as MAP\_RMS maps if processed 
as follows: $ weight map =  {F_A \over sqrt(drizzle_weight_map)}$ with 
$F_A$ a certain correction parameter to get pure, uncorrelated RMS noise. This value 
depends on the area over which the noise is determined.
\cite{casertano} estimated this in their appendix and came to this:
If the pixfrac of drizzle is p and the scale of the output 
pixels is s (0.5 half of the original pixels) then the ratio between 
uncorrelated noise and the drizzle map value is:

$$ sqrt(F_A) = ({s\over p}(1 -{1\over3} {s \over p}) \ , \ if \ s<p  $$

$$ sqrt(F_A) = 1 -{1\over3} s/p) \ , \ if \ s>p  $$


If you want to use this trick, please go over this bit in 
\cite{casertano} to be sure.

The resulting map can be used with WEIGH\_TYPE MAP\_RMS and thus takes in 
account all the weird thing drizzle did to the data (per pixel different 
contributions from pixels with maybe different exposures).

\chapter{SE parameters}

Below the settings of SE are listed with their function and the best way to use them in a crowded field. These values were obtained after using SE on a series of simulations of Sextans A (north pointing wf2) and the HDF. The success rate was measured by the number of objects classified as galaxies in the simulations actually found in the HDF with the same settings.

\begin{verbatim}

CATALOG_TYPE	ASCII	# This means no header with the values 
			# in different colums is printed. 
PARAMETERS_NAME	/net/bartoli/bartoli/data2/Programs/default2.param	
			# file with the names of the parameters to be 
			# printed in the catalogs.
DETECT_TYPE	CCD	# type of image (alt PHOTO)	
                
DETECT_MINAREA	16	# Minimum area of connected pixels in an object.
FILTER		Y	
FILTER_NAME	/net/bartoli/bartoli/data2/Programs/sextractor2.1.6/config/gauss_4.0_7x7.conv

DEBLEND_NTHRESH	32	# the number of thresholds the intensity 
			# range is devided up in. 32 is the most 
			# common number.
DEBLEND_MINCONT	0.005	# percentage of flux a separate peak must 
			# have to be considered a separate object.
CLEAN		Y	# Should objects close to bright object be 
			# removed from the catalogs? 	
CLEAN_PARAM	1.0	# measure for 'clean'ing. wish I knew what it ment...
PHOT_APERTURES	5,10,20,30	# Fixed apertures in pixels 
PHOT_AUTOPARAMS	2.5, 3.5	
SATUR_LEVEL	200000.	# saturation occurs at?
MAG_GAMMA	4.0	# Emulsion response slope
			# SE doesn't work without this for some reason
GAIN		7.0	# number of photons / counts ratio
			# if run over a counts per second image then the 
			# gain*exp time is used.
PIXEL_SCALE	0.05	# arcsec
SEEING_FWHM	.17	# 
STARNNW_NAME	/net/bartoli/bartoli/data2/Programs/sextractor2.1.6/config/default.nnw
BACK_SIZE	32	# size in pixels of the area used to estimate the 
			# background
BACK_FILTERSIZE	1	# type of filter used in the background 
BACKPHOTO_TYPE	LOCAL	# local estimate or an estimate for the entire 
			# image at once...	
BACKPHOTO_THICK	32	# 

#CHECKIMAGE_TYPE	APERTURES	# output image of SE.
#CHECKIMAGE_NAME	Roza_F555W_wf3.fits # name of that image.
CHECKIMAGE_TYPE	NONE	
MEMORY_OBJSTACK	10000	# memory parameters	
MEMORY_PIXSTACK	1500000	# defaults work best
MEMORY_BUFSIZE	1024		
#SCAN_ISOAPRATIO	0.6	
VERBOSE_TYPE	NORMAL		
\end{verbatim}


\bibliography{SE_Handbook.3} 
\bibliographystyle{astron}

\markboth{Index}{Index}
\printindex



\end{document}